\documentclass[a4paper]{article}

\usepackage{arxiv}
\usepackage[utf8]{inputenc} % allow utf-8 input
\usepackage[T1]{fontenc}    % use 8-bit T1 fonts
\usepackage{hyperref}       % hyperlinks
\usepackage{url}            % simple URL typesetting
\usepackage{booktabs}       % professional-quality tables
\usepackage{amsfonts}       % blackboard math symbols
\usepackage{nicefrac}       % compact symbols for 1/2, etc.
\usepackage[final,nopatch=footnote]{microtype}      % microtypography
\usepackage{lipsum}		% Can be removed after putting your text content
\usepackage{graphicx}
\usepackage{natbib}
\usepackage{float}
\usepackage{mathtools}
\usepackage{amsmath}
\usepackage{amssymb}
\usepackage{setspace}
\usepackage{hyperref,soul} 
\usepackage{threeparttable}
\usepackage{bbm}
\usepackage{tikz}
\usetikzlibrary{positioning, arrows.meta}
\newcommand{\bbeta}{\boldsymbol{\beta}}
\newcommand{\balpha}{\boldsymbol{\alpha}}
\newcommand{\btheta}{\boldsymbol{\theta}}
\newcommand{\bvartheta}{\boldsymbol{\vartheta}}
\newcommand{\bdelta}{\boldsymbol{\delta}}
\newcommand{\bGamma}{\boldsymbol{\Gamma}}
\newcommand{\bgamma}{\boldsymbol{\gamma}}
\newcommand{\bSigma}{\boldsymbol{\Sigma}}
\newcommand{\br}{\boldsymbol{r}}
\newcommand{\bs}{\boldsymbol{s}}
\newcommand{\bu}{\boldsymbol{u}}
\newcommand{\bx}{\boldsymbol{x}}

\newcommand{\cX}{\mathcal{X}}
\newcommand{\cS}{\mathcal{S}}
\newcommand{\cC}{\mathcal{C}}

\newcommand{\cT}{\mathcal{T}}
\newcommand{\by}{\boldsymbol{y}}

\newcommand{\prob}{\mathbb{P}}
\newcommand{\bQ}{\boldsymbol{Q}}
\newcommand{\const}{\mathrm{Const}}

\usepackage[varbb]{newpxmath}

\usepackage{amsmath}

\title{Bayesian copula-based modelling for multi-type spatio-temporal epidemic data}

\author{Matthew Adeoye\\
Department of Statistics\\
University of Warwick
\And Simon E.F. Spencer\\
Department of Statistics\\
University of Warwick
\And Xavier Didelot\\
School of Life Sciences and\\
Department of Statistics\\
University of Warwick 
}

\begin{document}
\maketitle

\begin{abstract}
    The study of infectious disease epidemiology for multi-type disease pathogens requires modelling techniques that account for the complex interactions existing between strains across geography and time.  In this paper, we propose a novel multi-type spatio-temporal infectious disease model to better support the understanding of these pathogens. We formulate a joint state-space for all epidemics arising for a given multi-type pathogen as well as biologically informed representations of how these epidemic states may interact. We introduce the use of several copula models to uncover the dependence structure of epidemics between strains. We develop a computationally efficient Markov chain Monte Carlo (MCMC) sampling scheme for all proposed models. We also provide robust model comparison techniques using bridge sampling and importance sampling to evaluate model evidence in high-dimensional space. We demonstrate the performance of our proposed models using simulated datasets, where simulated epidemics were successfully identified and associated parameters correctly inferred. The proposed models were also fitted to monthly multi-type incidence data on invasive meningococcal disease from 26 European countries. The accompanying software is freely available as a R package at \href{https://github.com/Matthewadeoye/MultiOutbreaks}{https://github.com/Matthewadeoye/MultiOutbreaks}.
\end{abstract}

\section{Introduction}

Infectious disease surveillance often includes typing of 
at least a subset of the reported infected cases. 
One of the oldest and most popular typing method is serotyping, 
in which antibodies are used to detect the presence of antigens
on the surface of the pathogen. Over a century after it was first
proposed \citep{neufeld1902ueber}, 
serotyping is still in widespread use for 
both bacterial pathogens such as 
\textit{Streptococcus pneumoniae} \citep{narciso2025streptococcus} 
or \textit{Neisseria meningitidis} \citep{rouphael2011neisseria}
and viral pathogens 
such as influenza \citep{nuttens2022evolution}
or foot-and-mouth disease \citep{jamal2013foot}.
There are many other typing methods depending on the disease, with
the most recent ones being based on molecular sequences, such as
multi-locus sequence typing (MLST) for bacteria \citep{maiden2006multilocus}, 
real-time polymerase chain reaction (PCR) for DNA viruses \citep{mackay2002real} and 
real-time reverse transcription PCR for RNA viruses \citep{bustin2005real}. 
In the remainder of this paper we consider that an infectious disease can be subdivided
into several types, irrespective of the exact typing method being used.

The most commonly used mathematical models of infectious disease epidemiology 
take a compartmental form \citep{brauer2008compartmental}. 
For example, the famous SIR model divides individuals in the 
population into
three compartments: susceptible, infected and removed \citep{kermack1927contribution}. 
Compartmental models can be extended in many ways, 
including to handle typing data, by duplicating
the compartments corresponding to infected individuals as many times as there are types
\citep{gog2002dynamics}. Additional parameters can then be used in the rates of transition
between compartments to represent the many types of interactions that may exist between types. 
In particular, infection with one type may result in immunity, at least for some time,
to re-infection with the same type, but may also result in cross-immunity, that is
partial or complete immunity to subsequent infection with other types 
\citep{castillo-chavezEpidemiologicalModelsAge1989,white1998cross,restifIntegratingLifeHistory2006}.
The possibility of superinfection with more than one type simultaneously may also be 
important to take into account for some diseases, resulting in additional compartments with
specific dynamic properties \citep{cohen2006beneficial,spicknall2013modeling}.

Here we focus on a different class of infectious disease model,
specifically designed to analyse spatio-temporal data, where a 
number of cases is observed for each spatial and temporal unit
\citep{knorr2003hierarchical,bakar2015sptimer,meyer2017spatio}.
In this context, the joint modelling of multiple types has not yet been
explored as for compartmental models. Consequently, analysis of 
multi-type data with currently existing spatio-temporal methods requires to either
analyse each type separately, or aggregate all types as a single disease.
Neither of these two approaches can reveal interactions between types,
even though this is often the most important aspect to understand
 the mechanisms of disease spread and how epidemics arise. 
For example, cross-immunity has a long history of being
studied using compartmental models 
\citep{castillo-chavezEpidemiologicalModelsAge1989,white1998cross,restifIntegratingLifeHistory2006}
but can not currently be studied using spatio-temporal models.
Interaction between types is also crucial to take 
into account when formulating control intervention strategies
for many infectious diseases,
including influenza \citep{ferguson2005strategies,andersson2012stochastic}, 
antibiotic resistant gonorrhoea \citep{fingerhuth2016antibiotic,whittles2017estimating} 
and COVID-19 \citep{walker2020impact,dyson2021possible}.

We take as a starting point a previously proposed framework for
Bayesian spatio-temporal modelling for infectious disease epidemiology analysis
\citep{knorr2003hierarchical,spencer2011detection,ADEOYE2026100879} 
and restructure it to enable the joint analysis of multiple types. 
We propose to model the joint dynamics of multiple disease types using  copulas 
\citep{nelsen2006introduction}, 
which has the desirable property of separating their interdependency from their marginal behaviour.
The resulting models are complex with many parameters, but we show
that Bayesian inference can still be performed at scale using a carefully
designed Metropolis adjusted Langevin algorithm \citep{girolami2011riemann}.
We evaluate the usefulness of this new inferential methodology 
using simulated datasets,
in which the correct model and parameter values are known so that 
the accuracy of inferred values can be assessed. We also showcase
the practical use of our methodology with an application to real data 
on the incidence of the main four serogroups of invasive meningococcal
disease across European countries \citep{ECDC_Atlas}.

\section{Methods}
\subsection{Model structure}
Let $y_{it}$ and $e_{it}$ denote the number of cases and the size of the population at risk at location $i$ at time $t$, respectively for $i=1,...,I$ and $t=1,...,T$. The observed case counts, $y_{it}$, conditional on the risks $\lambda_{it}$, are assumed to follow a Poisson distribution:
\begin{equation} \label{eq:model1}
y_{it} | \lambda_{it} \sim \text{Poisson}(e_{it}\lambda_{it}).
\end{equation}
We decompose the logarithmic risk of disease incidence into a temporal trend component ($r_t$), a seasonal component ($s_{t\bmod C}$), a spatial component ($u_i$), a spatio-temporal epidemic indicator ($x_{it}=1$ for the epidemic state and $x_{it}=0$ for the endemic state),  and an increased risk factor during epidemics ($\beta$): 
\begin{equation} \label{eq:model2}  
\log(\lambda_{it})= a + r_t + s_{t\bmod C} + u_i + x_{it}\beta.
\end{equation}
The epidemic indicator $x_{it}$ is ruled by a time-homogeneous Markov chain.
Note that this model corresponds to the model VII proposed by \cite{ADEOYE2026100879}.

\subsection{Multi-type disease pathogens}
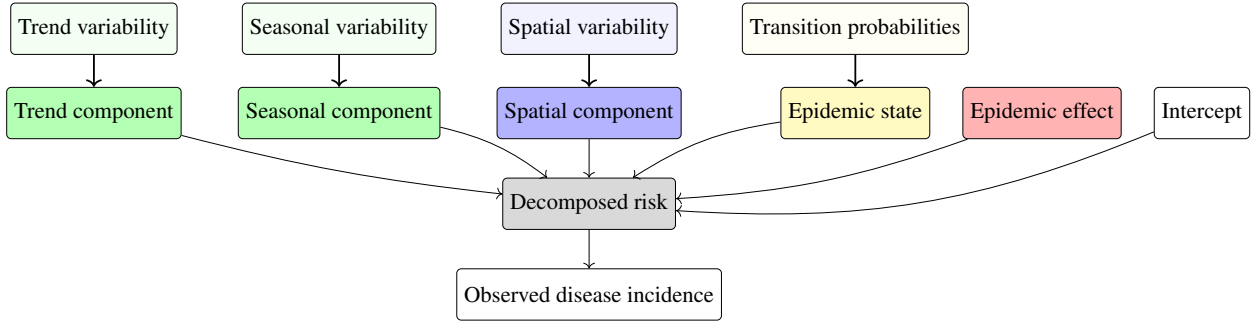
\begin{figure}[httb]
    \centering
    \resizebox{\textwidth}{!}{%
\begin{tikzpicture}[
    node distance = 1.5cm and 1.0cm,
    every node/.style = {
        draw,
        rectangle,
        rounded corners=2pt,
        minimum width=1.0cm,
        minimum height=0.8cm,
        align=center,
        font=\normalsize
    },
    every edge/.style = {->, thick, >=Stealth},
]
\node[draw=black, fill=green!5] (kr) {Trend variability};
\node[draw=black, fill=green!5] (ks) [right=of kr] {Seasonal variability};
\node[draw=black, fill=blue!5] (ku) [right=of ks] {Spatial variability};
\node[draw=black, fill=yellow!5] (G)  [right=of ku] {Transition probabilities};

\node[draw=black, fill=green!30] (r) [below=0.5cm of kr] {Trend component};
\node[draw=black, fill=green!30] (s) [below=0.5cm of ks] {Seasonal component};
\node[draw=black, fill=blue!30] (u) [below=0.5cm of ku] {Spatial component};
\node[draw=black, fill=yellow!30] (x) [below=0.5cm of G]  {Epidemic state};
\node[draw=black, fill=red!30] (b) [right=0.5cm of x]  {Epidemic effect};
\node[draw=black] (a) [right=0.5cm of b] {Intercept};

\node[draw=black, fill=gray!30] (L) [below=0.6cm of u] {Decomposed risk};
\node (y) [below=0.6cm of L] {Observed disease incidence};

\draw[->, thick] (kr) -- (r);
\draw[->, thick] (ks) -- (s);
\draw[->, thick] (ku) -- (u);
\draw[->, thick] (G)  -- (x);

\draw[->, bend left=13]  (a) to (L);
\draw[->, bend left=-4]  (r) to (L);
\draw[->, bend left=12]  (s) to (L);
\draw[->]                (u) to (L);
\draw[->, bend right=12] (x) to (L);
\draw[->, bend right=-8] (b) to (L);

\draw[->] (L) -- (y);
\end{tikzpicture}
}
 \caption{Directed acyclic graph (DAG) illustrating the relationships between all model components and the observed disease incidence.}
    \label{fig:DAG_model}
\end{figure}

Let us consider that there are $K \in \mathbbm{N}$ strains of the disease pathogen being studied, and we are interested in epidemics in all strains, both marginally and jointly. We extend the model in Equation \ref{eq:model1} as follows:
\begin{equation} \label{eq:model-multitype}
y_{itk} | \lambda_{itk} \sim \text{Poisson}(e_{it}\lambda_{itk}).
\end{equation}
The simplest formulation of the logarithmic risk $\lambda_{itk}$ in Equation \ref{eq:model-multitype} 
is to assume that there are $K$ independent epidemic terms for all $K$ types of disease pathogens observed:
\begin{equation} \label{eq:multitype-model}  
\log(\lambda_{itk})= a_k + r_t + s_{t\bmod C} + u_i + x_{itk} \beta_k.
\end{equation}
The difference with the previous model in Equation \ref{eq:model2} is that $a_k$ is the intercept for the $k$-th pathogen and $\beta_k$ is the increased risk factor associated with the $k$-th epidemic term ($x_{itk}$). A graphical illustration of the model is shown in Figure \ref{fig:DAG_model}.

\subsubsection{Spatio-temporal model for the sporadic cases}
The spatio-temporal model for the sporadic cases excludes the epidemic component in Equation \ref{eq:multitype-model}, where $a_k$ remains the strain-specific intercept. The trend, seasonal and spatial components are given priors from the family of intrinsic Gaussian Markov random fields \citep{rue2005Gaussian}, designed to induce smoothness in either space or time. The trend component $r_t$ is assumed to follow a smooth second-order random walk model given by:
\begin{equation} \label{eq:trendcomp}
r_t | r_{t-1}, r_{t-2} = 2r_{t-1} - r_{t-2} + \epsilon_t, 
\end{equation}
where $\epsilon_t \sim \mathcal{N}\left(0, \kappa_r^{-1}\right)$ and $\kappa_r$ is the precision parameter. Similarly, the seasonal component $s_{t\bmod C}$ is assumed to follow a cyclical first-order random walk written as:
\begin{equation} \label{eq:seasonalcomp}
s_c= s_{c-1} + \epsilon_c, 
\end{equation}
where $\epsilon_c \sim \mathcal{N}\left(0, \kappa_s^{-1}\right)$, $\kappa_s$ is the precision parameter, $c=2,\dots,C$, and then $s_1-s_C\sim \mathcal{N}\left(0, \kappa_s^{-1}\right)$ so that $s_1$ and $s_{C}$ are considered neighbouring components. Lastly, the spatial component $u_i$ is assumed to follow an intrinsic conditional autoregression (ICAR) model which acts as a spatial smoothing prior and is typically written as:
\begin{align} \label{eq:spatialcomp}
    u_{i}=\frac{1}{|n(i)|}\sum_{j \in n(i)}u_j+ \epsilon_i,
\end{align}
where $n(i)$ is the set of indices of locations that neighbour location $i\in\{1,\dots,I\}$, and $\epsilon_i\sim\mathcal{N}\left(0,\frac{1}{|n(i)|\kappa_u}\right)$, $\kappa_u$ is the precision parameter. These temporal and spatial components are assumed to be constant across all strains. Full details on these prior distributions are provided in Appendix Sections \ref{sec:trendAppendix}, \ref{sec:seasonalAppendix}, and \ref{sec:spatialAppendix}.

\subsubsection{Joint state-space formulation for the epidemic cases}
Let us now consider the joint state-space for epidemics in all $K$ strains.  
We define $\boldsymbol{x}_{it}=(x_{it1}, x_{it2}, \dots,x_{itK})\in \{0,1\}^K$
to be a Markov process over time $t$ with transition matrix $\bGamma$.
A first model is to consider that epidemics of the different strains happen
independently and according to the same Markov chain 
with 2x2 transition matrix $\bgamma$.
For example, if there are $K=2$ strains each ruled by 
$\bgamma =
 \begin{pmatrix}
  1-p & p\\ 
  q & 1-q \\
\end{pmatrix}$ then
%with stationary distribution $\boldsymbol{\delta}=(\delta_1,\delta_2)=\frac{1}{p+q}(q,p)$,
%\begin{align*}
%\gamma &=
% \begin{pmatrix}
%  1-p & p\\ 
%  q & 1-q \\
%\end{pmatrix} = \begin{pmatrix}
%  1-p & p\\ 
%  \frac{\delta_1}{\delta_2}p & 1-\frac{\delta_1}{\delta_2}p \\
%\end{pmatrix},
%\end{align*}
the resulting joint transition probability matrix is:
\begin{align}
\bGamma &=
 \bordermatrix{
 &(0,0)&(0,1)&(1,0)&(1,1)\cr
(0,0)&  (1-p)^2 & (1-p)p & p(1-p) & p^2\cr
(0,1)&  (1-p)q & (1-p)(1-q) & pq & p(1-q)\cr
(1,0)& q(1-p) & qp & (1-q)(1-p) & (1-q)p\cr
 (1,1)&q^2 & q(1-q) & (1-q)q & (1-q)^2\cr 
 }
 \label{eq:Gamma-indep-k2}
\end{align}
%in which the order of the rows and columns correspond to having no epidemic,
%an epidemic of strain 1 only, an epidemic of strain 2 only, and an epidemic
%of both strains, respectively.
To write down the general case $K\geq2$ we need to consider how the $i$-th state
of the joint model relates to the epidemic state for each of the $K$ strains.
We use the following encoding: the state $i\in \{0,...,2^K-1\}$ is such
that there is an epidemic of the $k$-th strain if the $k$-th bit of $i$ is one.
Extracting the $k$-th bit of $i$ is done via integer division and modulo:
$$\mathrm{bit}(i,k)=\left\lfloor \frac{i}{2^{k-1}}\right\rfloor \bmod 2$$
We can then write down the transition probability from state $i$ to state $j$ as:
\begin{align}\label{eq:Gamma-model1}
\Gamma_{i,j}=\prod^K_{k=1}\gamma_{\mathrm{bit}(i,k),\mathrm{bit}(j,k)} \ \ \ \ \text{for} \ \ i,j \in \{0,\dots,2^K-1\}.
\end{align}

An alternative model is to consider that epidemics of the different strains happen independently, but each is ruled by its own unique transition probability matrix, $\bgamma^{(k)}$. In this case the joint transition matrix is:
\begin{align}\label{eq:Gamma-model2}
\Gamma_{i,j}=\prod^K_{k=1}\gamma^{(k)}_{\mathrm{bit}(i,k),\mathrm{bit}(j,k)} \ \ \ \ \text{for} \ \ i,j \in \{0,\dots,2^K-1\}.
\end{align}

\subsubsection{Copulas}
Sklar's theorem states that any multivariate cumulative distribution function (CDF), 
$F(\bx)=\prob(X_1\leq x_1,\dots,X_K\leq x_K)$ %, $\bx \in \mathbb{R}^K$ 
can be expressed in terms of its
univariate marginal distributions $F_i(x_i)=\prob(X_i \leq x_i)$ and a copula function 
$\cC:[0,1]^K\rightarrow[0,1]$ such that 
$F(\bx)=\cC(F_1(x_1),\dots,F_K(x_K))$
\citep{sklar1959fonctions}. A copula therefore captures the dependence structure of a vector of random variables, separately from their marginal behaviours. There are many copula families in the literature \citep{hofert2018elements,joe2014dependence,nelsen2006introduction}, each modelling specific dependence structures that allow flexible representation of complex multivariate relationships. In this paper, we consider the Frank copula and the Gaussian factor copula model. 

A $K$-dimensional Frank copula, for any $\bu\in[0,1]^K$, is defined as:
\begin{align}
    \cC_\psi\bigl(\bu\bigr)=-\psi^{-1}\log{\left\{1-\frac{\prod_{k=1}^K(1-\exp{(-\psi u_k)})}{(1-\exp{(-\psi))^{K-1}}}\right\}}.
    \label{eq:frank}
\end{align}
The Frank copula has a single parameter with support $\psi \in \mathbb{R}\setminus\{0\}$ for the bivariate case, and $\psi\in \mathbb{R}^+$ for any dimension $K\geq3$, and this parameter characterizes the dependence structure between the variables. The multivariate Frank copula with dimension $K\geq3$ can only model positive dependence structure due to its inability to satisfy the monotonicity constraint on its generator for negative dependence structures \citep{hofert2018elements}. In contrast, the bivariate Frank copula satisfies the monotonicity constraint for both positive and negative dependence structures and has the following interpretation: (i) $\psi\rightarrow0$ corresponds to independence, (ii) $\psi>0$ corresponds to a positive dependence structure, and (iii) $\psi<0$ corresponds to a negative dependence structure.

A $K$-dimensional Gaussian copula is the copula obtained via Sklar's theorem from the multivariate Gaussian distribution 
$\mathcal{N}_K(\boldsymbol{0},\bSigma)$ with correlation matrix $\bSigma$. Note that $\bSigma$ is both the correlation and covariance matrix here since the random variables ($x_1,\dots,x_K$) are standard normal variates. If $\Phi_{\bSigma}$ denotes the multivariate Gaussian cumulative distribution function, the copula $\mathcal{C}_{\bSigma}$ is given, for any $\bu\in[0,1]^K$ by
\begin{align}
    \cC_{\bSigma}\bigl(\bu\bigr)
&= \Phi_{\bSigma}\bigl(\Phi^{-1}(u_1),\dots,\Phi^{-1}(u_K)\bigr) \notag \\
&= \int_{-\infty}^{\Phi^{-1}(u_K)}\!\!\cdots 
\int_{-\infty}^{\Phi^{-1}(u_1)}
\frac{\exp\!\left(-\frac12\bx^\top\bSigma^{-1}\bx\right)}
{(2\pi)^{K/2}\sqrt{\det \bSigma}}
\,\mathrm{d}x_1 \dots \mathrm{d}x_K,
\end{align}
where $\Phi^{-1}$ denotes the quantile function of $\mathcal{N}(0,1)$. However, the $K$-dimensional Gaussian copula typically requires evaluating a $K$-dimensional integral, which is computationally intensive as $K$ increases. We consider the factor copula model proposed to capture tail dependence and asymmetries in multivariate data while offering parsimony and significant computational advantages over the multivariate Gaussian copula \citep{krupskii2013factor}. The Gaussian factor copula model attempts to explain the dependence structure in the observed variables by one or more latent variables. For example, the $j$-factor copula assumes $\bu=(u_1,\dots,u_K)$ to be conditionally independent given $j$ latent variables $V_1,\dots,V_j$, where the latent variables are assumed to be independently and identically distributed as Uniform$(0,1)$. Let the conditional CDF and copula of the $k$th variable given $V_1,\dots,V_j$ be denoted as $F_{k|V_1,\dots,V_j}$ and $C_{k|V_1,\dots,V_j}$ respectively. Then
\begin{align*}
\cC\bigl(\bu\bigr)&=\int_0^1\dots\int_0^1\prod_{k=1}^KF_{k|V_1,\dots,V_j}(u_k|v_1,\dots,v_j)\:\mathrm{d}v_1\dots\mathrm{d}v_j. 
\end{align*}
Similarly, the $1-$factor copula model conditions each marginal distribution on a common latent variable $V_1$ and is written as:
\begin{align*}
\cC\bigl(u_1,\dots,u_K\bigr)&=\int_{0}^1\prod_{k=1}^KF_{k|V_1}(u_k|v_1)\:\mathrm{d}v_1=\int_{0}^1\prod_{k=1}^K\cC_{k|V_1}(u_k|v_1)\:\mathrm{d}v_1.
\end{align*}
Suppose $u_k=\Phi(x_k)$, then the $1-$factor copula model using a bivariate Gaussian  copula with correlation parameter $\xi_{k1}$ for linking the $k$-th strain with the latent variable $V_1$, is written as:
\begin{align}\label{eq:factor}
    \cC\bigl(\Phi(x_1),\dots,\Phi(x_K)\bigr)&=\int_0^1\prod_{k=1}^K\Phi\left(\frac{x_k-\xi_{k1}\Phi^{-1}(v_1)}{\sqrt{1-\xi_{k1}^2}}\right)\mathrm{d}v_1=\int_{-\infty}^\infty\left\{\prod_{k=1}^K\Phi\left(\frac{x_k-\xi_{k1}w}{\sqrt{1-\xi_{k1}^2}}\right)\right\}\phi(w)\:\mathrm{d}w,
\end{align}
where $\Phi$ and $\phi$ represent the Gaussian cumulative distribution function and probability density function, respectively. The correlation parameters $\{\xi_{k1}:k=1,\dots,K\}$ between each disease strain and the latent variable are referred to as factor loadings. Thus, the dependence structures between the variables ($x_1,\dots,x_K$) are derived from factor loadings, such as Corr$(x_k,x_l)=\xi_{k1}\times\xi_{l1}$.

\subsubsection{Epidemic dependence modelling via copulas}

Through the use of copulas, we can relax the assumption of independence in transitions between strains. 
Let us once again consider first the case $K=2$. Let $\cC(u,v)$ denote an arbitrary bivariate copula where $u$ and $v$ denote the marginal probabilities that strains 1 and 2 transit to the epidemic state 1 in the next step, respectively. The four joint probabilities follow from Sklar’s theorem and the inclusion-exclusion principle \citep{nelsen2006introduction,joe2014dependence}:
\begin{align*}
    \prob(X_{t}=(0,0)|X_{t-1})&=1-u-v+\cC(u,v),\\
    \prob(X_{t}=(0,1)|X_{t-1})&=v-\cC(u,v),\\
    \prob(X_{t}=(1,0)|X_{t-1})&=u-\cC(u,v),\\
    \prob(X_{t}=(1,1)|X_{t-1})&=\cC(u,v).
\end{align*}
Assuming the two strains have the same transition matrix 
$\bgamma =
 \begin{pmatrix}
  1-p & p\\ 
  q & 1-q \\
\end{pmatrix}$, then the elements of the joint transition matrix are
 obtained by replacing $(u,v)$ with $(p,p)$ for the transitions from state $(0,0)$;
with $(p,1-q)$ for the transitions from state $(0,1)$; with $(1-q,p)$ for the transitions from state $(1,0)$; and with $(1-q,1-q)$ for the transitions from state $(1,1)$.
These calculations are detailed in the Appendix \ref{sec:copulaAppendixK2}, as well as for the case $K=3$ in Appendix \ref{sec:copulaAppendixK3}. 
Gathering these terms into a matrix gives the transition matrix:
\begin{align}\label{eq:Gamma-dep-k2}
\bGamma &=
\bordermatrix{
&(0,0)&(0,1)&(1,0)&(1,1)\cr
(0,0)&  1-2p+ \cC(p,p) & p- \cC(p,p) & p- \cC(p,p) & \cC(p,p)\cr
(0,1)&  q-p+\cC(p,1-q) & 1-q-\cC(p,1-q) & p-\cC(p,1-q) & \cC(p,1-q)\cr
(1,0)& q-p+\cC(1-q,p) & p-\cC(1-q,p) & 1-q-\cC(1-q,p) & \cC(1-q,p)\cr
(1,1)& 2q-1+\cC(1-q,1-q) & 1-q-\cC(1-q,1-q) & 1-q-\cC(1-q,1-q) & \cC(1-q,1-q)\cr
}. 
\end{align}

Note that the transition matrix  
for the independent model (Equation \ref{eq:Gamma-indep-k2})
is a special
case of the dependent model (Equation \ref{eq:Gamma-dep-k2})
if we use the 
independence Copula $\cC(u,v)=uv$.
Generalization to the case $K\geq 2$ follows directly from the inclusion-exclusion principle and requires  a $K$-dimensional multivariate copula, $\cC$.
%, and when applied to a subset of marginal probabilities with fewer than $K$ arguments, it implicitly uses the marginal copula over those dimensions by setting the remaining arguments to 1 \citep{nelsen2006introduction}. 
%Note also that for consistency, $\cC(u) = u$ for a single argument and $\cC(\emptyset) = 1$ for the empty set.
Let $\cS = \{k: \mathrm{bit}(j,k)=1\}$ be the set of strains for which there are epidemics in state $j$, and let $\cS^\complement = \{k: \mathrm{bit}(j,k)=0\}$ be its complement. %Denote by  $\cP\left(\cS^\complement\right)$ the power set of $\cS^{\complement}$. 

If all strains have the same transition probability matrix $\gamma$ then each element of the copula-dependent joint transition probability matrix is given by: 
\begin{align}\label{eq:Gamma-copula1}
     \Gamma_{i,j} &= \sum_{\cT\subseteq \cS^\complement}(-1)^{|\cT|}\cC\left(\left\{(\gamma_{\mathrm{bit}(i,k),1})^{\mathbb{1}_{(k \in \cS \cup \cT)}}: k=1,...,K\right\}\right).
\end{align}

If on the other hand each strain has its own transition probability matrix
$\bgamma^{(k)}$ then:
\begin{align}\label{eq:Gamma-copula2}
     \Gamma_{i,j} &= \sum_{\cT\subseteq \cS^\complement}(-1)^{|\cT|}\cC\left(\left\{(\gamma^{(k)}_{\mathrm{bit}(i,k),1})^{\mathbb{1}_{(k \in \cS \cup \cT)}}: k=1,...,K\right\}\right).
\end{align}

Finally, we can define the general-dependent model to be any Markov matrix $\bGamma$ with rows adding up to one:
\begin{align}\label{eq:Gamma-general}
\bGamma =
 \begin{pmatrix}
  \gamma_{00\dots0,00\dots0} & \gamma_{00\dots0,00\dots1} & \dots & \gamma_{00\dots0,11\dots1}\\ 
   \gamma_{00\dots1,00\dots0} & \gamma_{00\dots1,00\dots1} & \dots & \gamma_{00\dots1,11\dots1} \\ 
 \vdots & \vdots & \ddots & \vdots \\ 
 \gamma_{11\dots1,00\dots0} & \gamma_{11\dots1,00\dots1} & \dots & \gamma_{11\dots1,11\dots1}\\ 
\end{pmatrix}\mathrm{~with~}\boldsymbol{\Gamma 1}^\top=\boldsymbol{1}^\top
\end{align}

In this paper we compare models using the independent copula, Frank copula, Gaussian factor copula and the general-dependent model. The copula models can be considered with shared or separate strain transition probabilities. The properties of the models that we have described are summarised in Table \ref{tab:AllModels}.

\begin{table}[H]
\centering
    \begin{threeparttable}
    \caption{\textbf{Summary of models used in this study}    \label{tab:AllModels}}
\begin{tabular}{ c c c c}
\hline
 Model type & Marginal transition & Joint transition  & Number of transition parameters \\
 \hline
 No epidemic & Nil & Nil & $0$ \\
 Independent 1 & $\bgamma$ & Equation \ref{eq:Gamma-model1}
 %$\Gamma_{a,b}=\prod^K_{k=1}\gamma_{\mathrm{bit}(a,k),\mathrm{bit}(b,k)}$ 
 & $2$ \\
% \hline
 Independent 2 & $\bgamma^{(k)}$ & Equation \ref{eq:Gamma-model2}
 %$\Gamma_{a,b}=\prod^K_{k=1}\gamma^{(k)}_{\mathrm{bit}(a,k),\mathrm{bit}(b,k)}$ 
 & $2K$ \\ 
  Frank copula 1 & $\bgamma$ &  Equations \ref{eq:frank} and \ref{eq:Gamma-copula1}
 %$\Gamma_{a,b} = \sum_{\cT\subseteq \cS^\complement}(-1)^{|\cT|}\cC\left(\left\{\gamma_{\mathrm{bit}(a,k),\mathrm{bit}(b,k)}: k=1,\dots,K\right\}_{\cS \cup \cT}\right)$  
 & $2+1$ \\ 
% \hline
  Frank copula 2 & $\bgamma^{(k)}$ &  Equations \ref{eq:frank} and \ref{eq:Gamma-copula2}
  %$\Gamma_{a,b} = \sum_{\cT\subseteq \cS^\complement}(-1)^{|\cT|}\cC\left(\left\{\gamma^{(k)}_{\mathrm{bit}(a,k),\mathrm{bit}(b,k)}: k=1,\dots,K\right\}_{\cS \cup \cT}\right)$  
  & $2K+1$ \\ 
%  \hline
 Gaussian factor copula 1 & $\bgamma$ &  Equations \ref{eq:factor} and \ref{eq:Gamma-copula1}
 %$\Gamma_{a,b} = \sum_{\cT\subseteq \cS^\complement}(-1)^{|\cT|}\cC\left(\left\{\gamma_{\mathrm{bit}(a,k),\mathrm{bit}(b,k)}: k=1,\dots,K\right\}_{\cS \cup \cT}\right)$  
 & $2+ K$ \\ 
% \hline
Gaussian  factor copula 2 & $\bgamma^{(k)}$ &  Equations \ref{eq:factor} and \ref{eq:Gamma-copula2}
  %$\Gamma_{a,b} = \sum_{\cT\subseteq \cS^\complement}(-1)^{|\cT|}\cC\left(\left\{\gamma^{(k)}_{\mathrm{bit}(a,k),\mathrm{bit}(b,k)}: k=1,\dots,K\right\}_{\cS \cup \cT}\right)$  
  & $2K+K$ \\ 
  %  \hline
 General-dependent & Nil & Equation \ref{eq:Gamma-general}
  & $2^K(2^K-1)$ \\ 
 \hline
\end{tabular}
    \end{threeparttable}
\end{table}

\subsection{Inference}
Let the parameters be denoted $\boldsymbol{\theta}=(\boldsymbol{r}_{1:T},\boldsymbol{s}_{1:C},\boldsymbol{u}_{1:I}, \kappa_r, \kappa_s, \kappa_u, \boldsymbol{a}_{1:K},\boldsymbol{\beta}_{1:K}, \bGamma)$, then the target posterior distribution can be decomposed using Bayes' theorem as:
\begin{align}
\label{eq:posterior}
\prob(\boldsymbol{\theta}|\boldsymbol{y}_{1:I,1:T, 1:K}) &\propto \prob(\by_{1:I,1:T,1:K}|\boldsymbol{\theta})\times \prob(\boldsymbol{\theta})\nonumber\\
&\propto \prob(\boldsymbol{\theta})\prod_{i=1}^I \prob(\by_{i,1:T,1:K}|\btheta)
\end{align}
using conditional independence between locations given $\bu$. %Moreover, $\{y_{itk}:k=1,\dots,K\}$ are also conditionally %independent given the joint Markov chain, $\bX_{it}$. 
%The likelihood terms can be decomposed as
%\begin{align*}
%%\label{eq:posterior}
%\prob(\by_{i,1:T,1:K}|\btheta)&= \sum_{\bX_{i,1:T}\in\{0,1\}^{KT}} \prob(\boldsymbol{y}_{i,1:T,1:K}, \bX_{i,1:T}|\boldsymbol{\theta}).
%\end{align*}

\subsubsection{Likelihood via forward filtering}
\label{sec:forward filtering}
Define the joint state space, $\cX=\{0,1\}^K$ for the joint Markov chain $\bx_{i,1:T}$, and let $\boldsymbol{\theta}_i=(\boldsymbol{r},\boldsymbol{s},u_i, \boldsymbol{a}, \boldsymbol{\beta}, \bGamma)$. We assume that the joint Markov chain starts from its stationary distribution, $\bdelta$. Given the full joint transition probability matrix, $\bGamma$, the stationary distribution is obtained by solving the linear system for $\bdelta$ such that $\bdelta = \bdelta\bGamma$ and $\bdelta\boldsymbol{1}^\top=1$ are both satisfied.
The forward filtering algorithm \citep{zucchini2009hidden} involves defining a sequence of vectors $\boldsymbol{\alpha}_{i,t}$ with length $|\cX|$, as follows:
\begin{align} \label{eq:forward vector}
 \alpha_{i,t}(\bx_{i,t}) &=\prob(y_{i,1:t,1:K},\bx_{i,t}|\boldsymbol{\theta}_i),   
\end{align}
for each $\bx_{i,t}\in \cX$. 
For the initial timepoint we have:
\begin{align*}
\alpha_{i,1}(\bx_{i,1}) &=\prob(y_{i,1,1:K},\bx_{i,1}|\boldsymbol{\theta}_i) \\
&=\prob(y_{i,1,1:K}|\bx_{i,1},\boldsymbol{\theta}_i)\prob(\bx_{i,1}|\boldsymbol{\theta}_i) \\
&=\left(\prod_{k=1}^K\prob(y_{i,1,k}|\bx_{i,1},\boldsymbol{\theta}_i)\right)\delta_{\bx_{i,1}}.
\end{align*}

For $t>1$ these forward vectors satisfy the recursion:
\begin{align*}
\alpha_{i,t}(\bx_{i,t}) &=\sum_{\bx_{i,t-1}\in \cX}\prob(y_{i,1:t,1:K},\bx_{i,t},\bx_{i,t-1}|\boldsymbol{\theta}_i), \\
&=\sum_{\bx_{i,t-1}\in \cX}
\prob(y_{i,t,1:K}|\bx_{i,t},y_{i,1:t-1,1:K},\bx_{i,t-1},\btheta_i)
\prob(\bx_{i,t}|y_{i,1:t-1,1:K},\bx_{i,t-1},\btheta_i)
\prob(y_{i,1:t-1,1:K},\bx_{i,t-1}|\btheta_i)\\
 &=\sum_{\bx_{i,t-1}\in \cX}
\left(\prod_{k=1}^K \prob(y_{i,t,k}|\bx_{i,t},\boldsymbol{\theta}_i)\right)
 \prob(\bx_{i,t}|\bx_{i,t-1},\boldsymbol{\theta}_i)
 \alpha_{i,t-1}
 (\bx_{i,t-1})
\end{align*}
The likelihood terms in Equation \ref{eq:posterior} can then be calculated as:
\begin{equation*}
\prob(\by_{i,1:T,1:K}|\btheta) =\sum_{\bx_{i,T}\in \cX}\alpha_{i,T}(\bx_{i,T}).
\end{equation*}

\subsubsection{Marginal probability of an epidemic}
\label{sec:outBprob}
The epidemic indicators in the spatio-temporal component, $\bx_{i,t}=(x_{i,t,1},\dots,x_{i,t,K})$, represent when and where an epidemic is occurring jointly across strains, where $x_{i,t,k}=0$ indicates the endemic state and $x_{i,t,k}=1$ indicates the epidemic state. Hence, the probability of an epidemic can be obtained by local decoding using the backward part of the forward filtering backward sampling algorithm \citep{zucchini2009hidden}.
We are interested in the conditional probability $\prob({\bx_{i,t}}|\boldsymbol{y}_{i,1:T,1:K}, \boldsymbol{\theta}_i)$, for each $t=1,\dots,T$.

We define a sequence of $T$ backward vectors, each with length $|\cX|=2^K$:
\begin{equation} \label{eq:backward vector}    
\beta_{i,t}(\bx_{i,t}) =\prob(\by_{i,(t+1):T,1:K}|\bx_{i,t},\boldsymbol{\theta}_i).
\end{equation}
In particular we have:
\begin{equation*}
\beta_{i,T}(\bx_{i,T}) =\prob(\Omega|\bx_{i,T},\boldsymbol{\theta}_i) = 1.
\end{equation*}

For $t=T-1,\dots,1$, the backward vectors satisfy the recursion:
\begin{align*}
\beta_{i,t-1}(\bx_{i,t-1}) &=\prob(\by_{i,t:T,1:K}|\bx_{i,t-1},\boldsymbol{\theta}_i)\\
&= \sum_{\bx_{i,t}\in \cX}
\prob(\by_{i,t:T,1:K}|\bx_{i,t},\boldsymbol{\theta}_i)
\prob(\bx_{i,t}|\bx_{i,t-1},\boldsymbol{\theta}_i)\\
 &=\sum_{\bx_{i,t}\in \cX}
 \prob(\by_{i,(t+1):T,1:K}|\bx_{i,t},\boldsymbol{\theta}_i)
 \prob(\by_{i,t,1:K}|\bx_{i,t},\boldsymbol{\theta}_i)
 \prob(\bx_{i,t}|\bx_{i,t-1},\boldsymbol{\theta}_i)
  \\
 &=\sum_{\bx_{i,t}\in \cX}
\beta_{i,t}(\bx_{i,t})\left(\prod_{k=1}^K
 \prob(y_{i,t,k}|\bx_{i,t}, \boldsymbol{\theta}_i)\right)
 \prob(\bx_{i,t}|\bx_{i,t-1},\boldsymbol{\theta}_i).
\end{align*}

With the forward $\balpha_t$ and backward $\bbeta_t$ vectors defined as Equations \ref{eq:forward vector} and \ref{eq:backward vector} we can compute the marginal probability of an epidemic for each location and date: 
\begin{align*}
\prob({\bx_{i,t}}|\boldsymbol{y}_{i,1:T,1:K},\boldsymbol{\theta}_i)
&=\frac{\prob(\by_{i,1:T,1:K},\bx_{i,t} |\boldsymbol{\theta}_i)}
{\prob(\by_{i,1:T,1:K}|\boldsymbol{\theta}_i)}\\
&=\frac{\prob(\by_{i,1:t,1:K},\bx_{i,t} |\boldsymbol{\theta}_i)\prob(\by_{i,(t+1):T,1:K} |\bx_{i,t},\by_{i,1:t},\boldsymbol{\theta}_i)}{\prob(\by_{i,1:T,1:K}|\boldsymbol{\theta}_i)}\\
&=\frac{\alpha_{i,t}(\bx_{i,t}) \beta_{i,t}(\bx_{i,t})}{\sum_{\bx_{i,T}\in \cX}\alpha_{i,T}(\bx_{i,T})}. 
\end{align*}

\subsubsection{Missing and untyped data}
\label{sec:MissingData}
In many real-world applications, statistical inference is often complicated by the presence of missing data. The methodology presented in this paper and its accompanying R package (\href{https://github.com/Matthewadeoye/MultiOutbreaks}{MultiOutbreaks}) deal with scenarios of missing data by replacing the corresponding emission density in Sections \ref{sec:forward filtering} and \ref{sec:outBprob} with 1 for all states when data are missing. This is implemented under the assumption that the data are missing at random (MAR); see Section 2.3.4 in \citep{zucchini2009hidden} and \citep{yeh2012intermittent}. Similarly, there can be rare occasions of untyped data (that is, the overall incidence count is available, but no information on how it is distributed between strains). This is dealt with by summing the risks across all strains to evaluate the likelihood of the untyped incidence count, as done in previous studies \citep{benschop2021still}. 

\subsubsection{Priors}
Here we provide details and justifications for all prior distributions used for inference in the subsequent sections. For the transition probabilities in the independent and copula-dependent models, a Beta(1, 11) was chosen for the transition probability from the endemic to the epidemic state ($p$) to express a relatively low belief in such transitions. This prior has a monotonically decreasing density with one epidemic a year on average per strain, so that the model is conservatively using the epidemic state to prevent false positives. A Beta(6, 6) was chosen for the transition probability from the epidemic to the endemic state ($q$) to express a 0.5 probability of such transitions on average and therefore an average epidemic period of 2 months. For the general-dependent model, we chose a Dirichlet prior distribution $\text{Dir}\left(\nu_{i,0},\dots,\nu_{i,2^{K}-1}\right)$, $i\in \{0,\dots, 2^K-1\}$, for each simplex in the full transition probability matrix (Equation \ref{eq:Gamma-general}) based on the marginal priors on $p$ and $q$ from the copula models. The parameters of the Dirichlet are derived by matching the expectation of the independent models to the general-dependent model,  such that $\left(\nu_{i,0},\dots,\nu_{i,2^{K}-1}\right)=\frac{1}{12}\times\left(\mathbb{E}[\Gamma_{i,0}],\dots,\mathbb{E}[\Gamma_{i,2^{K}-1}]\right)$, with $\Gamma_{i,j}$ as defined in Equation \ref{eq:Gamma-model1}. In the Frank copula with $K=2$, a Normal(0, 100) prior was chosen for $\psi$, and for $K>2$, an Exponential(0.5) prior was chosen. In the Gaussian factor copula, we placed a flat Uniform(--1, 1) prior on the factor loadings. The prior distribution for each strain-specific epidemic effect ($\beta_k$) is a Gamma(2, 2) distribution which prevents label switching issues and ensures that the epidemic periods correspond to periods of strictly increased risk. These priors also facilitate a fair model comparison between all models in Table \ref{tab:AllModels}, see \cite{johnson2010use}. An informative prior Gamma$\left(0.01, \frac{0.01}{\exp{(-15)}}\right)$ was assumed for the transformed strain-specific intercept parameters $\phi_k=\exp{(a_k)}$ to allow a Gibbs-inspired proposal distribution during MCMC sampling. Following previous work \citep{ADEOYE2026100879}, the trend, seasonal and spatial components have prior densities as specified in Equations \ref{eq:trend prior}, \ref{eq:seasonal prior}, and \ref{eq:spatial prior} respectively. These three priors are defined in terms of the precision parameters $\kappa_r$, $\kappa_s$ and $\kappa_u$ for the trend component, seasonal component and spatial component, respectively. The hyper-priors of these three hyper-parameters are Gamma(1, 0.0001), Gamma(1, 0.001) and Gamma(1, 0.01), respectively.

\subsubsection{Monte Carlo sampling from the posterior}
\label{sec:MCMCImp}
The target posterior of our model is a high-dimensional probability density with strongly correlated temporal components. Gradient-based Markov chain Monte Carlo (MCMC) methods such as Hamiltonian Monte Carlo (HMC) \citep{duane1987hybrid,neal2011mcmc} or Metropolis adjusted Langevin algorithm (MALA) \citep{roberts1996exponential,roberts1998optimal} are known to have significant success in sampling efficiently from these kinds of probability densities, compared to traditional MCMC sampling methods. This is due to their ability to initiate ambitious moves within the state space whilst maintaining reasonable acceptance rates through the use of local problem-specific information. 

If $\btheta \in \mathbb{R}^D$ with density $p(\btheta)$ and log-density $\mathcal{L}(\btheta)=\text{log} \ p(\btheta)$, then MALA uses a discrete approximation to the Langevin diffusion which satisfies the following stochastic differential equation:
\begin{align}
    \text{d}\btheta(t)=\frac{1}{2}\nabla_{\btheta}\mathcal{L}(\btheta(t))\text{d}t + \text{d}\boldsymbol{B}(t),
\end{align}
where $\boldsymbol{B}(t)$ is a $D$-dimensional Brownian motion. The first-order Euler–Maruyama discretization of the continuous-time process yields a tractable discrete-time approximation of the form:
\begin{align}
    \btheta^*=\btheta^n + \frac{\varepsilon^2}{2} \nabla_{\btheta}\mathcal{L}(\btheta^n)+\varepsilon\boldsymbol{z}^n
\end{align}
where $\boldsymbol{z}\sim \mathcal{N}(\boldsymbol{0},\boldsymbol{\mathrm{I}})$ and $\varepsilon$ is the step size. The Metropolis accept-reject step \citep{metropolis1953equation,hastings1970monte} is then incorporated to correct for the error introduced by the step size $\varepsilon$, thus guaranteeing convergence to the correct invariant measure.

The gradient-based MCMC method described above converges to the desired stationary distribution more rapidly than the popular random-walk MCMC algorithm. However, MALA can still face shortcomings when sampling high-dimensional targets with strongly correlated parameters such as the second-order random walk used as the trend component in our model (Equation \ref{eq:trendcomp}). To address these shortcomings, a more efficient proposal is derived by noting that the space of parameterized probability density functions is endowed with a natural Riemann geometry \citep{girolami2011riemann}. The Riemannian manifold MALA (MMALA) proceeds by defining the Langevin diffusion having an invariant measure $p(\btheta)$, $\btheta \in \mathbb{R}^D$ directly on the Riemann manifold with arbitrary metric tensor $\boldsymbol{G}(\btheta)$. It was deduced that for a manifold with constant curvature, the position-specific preconditioned MALA proposal reduces to
\begin{align}
    \btheta^*=\btheta^n + \frac{\varepsilon^2}{2} \boldsymbol{G}^{-1}(\btheta^n) \nabla_{\btheta}\mathcal{L}(\btheta^n)+\varepsilon\sqrt{\boldsymbol{G}^{-1}(\btheta^n)}\boldsymbol{z}^n,
\end{align}
with the Metropolis accept-reject step to ensure convergence to the target measure.
We implement MMALA to update the temporal and spatial components of the model ($\br, \bs, \text{and} \ \bu$). The precision parameters ($\kappa_r$, $\kappa_s$, and $\kappa_u$) were sampled from their full conditional distributions, exploiting the conjugacy between their hyper-priors and the likelihood of the temporal and spatial components. The strain-specific intercepts ($a_k$) were sampled using Gibbs-inspired moves as described in Section \ref{sec:Gibbs-inspired}. The strain-specific regression parameters were each updated using simple Metropolis random walk moves with the proposal $\beta_{n+1}\sim\mathcal{N}(\beta_n, \sigma_\beta^2)$. The transition probabilities in the off-diagonals of $\bgamma$ in each of the independent and copula models were sampled using simple random walk moves. The copula parameter of the Frank copula models ($\psi$) is also updated using simple random walk moves. The copula parameters in the Gaussian factor copula model $\xi_{k,1}\in(-1,1)$ were each transformed to $z_k\in\mathbb{R}$ using a probit-type transformation $z_k=\Phi^{-1}\left(\frac{\xi_{k,1}+1}{2}\right)$ and then updated using simple Metropolis random walk moves as previously described, with the Jacobian correction $2\phi(z)$, where $\Phi^{-1}$ and $\phi$ represent the inverse cumulative distribution function and the probability density function of a standard Gaussian distribution, respectively. We impose an identifiability constraint by fixing $\xi_{1,1}\approx1$ since the factor model is invariant under global sign swapping, rotation, or scaling \citep{merkle2021efficient}. All simple random walk updates were implemented together with a Robbins-Monro adaptive tuning of the proposal standard deviations using the optimal high-dimensional target acceptance rate of 0.234 \citep{robbins1951stochastic}. Each row of the transition probability matrix $\bGamma$ in the general-dependent model was updated using Adaptive Dirichlet random-walk moves \cite{benschop2021still,spencer2021estimating} with a Dirichlet proposal $\bGamma_{n+1}^{(i)} \sim \text{Dir}\left(\nu_{i,0},\dots,\nu_{i,2^{K}-1} + h_n^{(i)}\bGamma_{n}^{(i)}\right)$, $i\in \{0, \dots, 2^K-1\}$, where $\left(h_n^{(i)}\right)$ is used to adaptively tune the proposal to stabilize the acceptance rate at roughly 25\%. After initializing log$\left(h_1^{(i)}\right)=0$, the subsequent tuning for log$\left(h_n^{(i)}\right)$ is given by log$(h_{n}^{(i)})=\text{log}\left(h_{n-1}^{(i)}\right)-3$ if a proposed move is accepted or log$\left(h_{n}^{(i)}\right)=\text{log}\left(h_{n-1}^{(i)}\right)+1$ if a proposed move is rejected. Finally, we use an adaptive Metropolis-Hastings random walk move with a multivariate Gaussian proposal and Robbins-Monro tuning \citep{robbins1951stochastic,vihola2011stability,spencer2021accelerating} for an additional joint update of the strain-specific intercepts and regression parameters, transition probabilities (with logit transformation), and copula parameters (with probit-type transformation for the factor loadings) with the appropriate Jacobian corrections included. The second option provided in our \textit{R} package is the dynamic HMC sampler via \textit{cmdstanr}, an \textit{R} interface to Stan \citep{carpenter2017stan}. This implementation offers the option to accelerate computations through a graphics processing unit (GPU) when available.

\subsubsection{Gibbs-inspired update for strain-specific intercepts}\label{sec:Gibbs-inspired}
The construction of an efficient proposal distribution is fundamental to the performance of a MCMC sampling algorithm. Here, we derive an approximate Gibbs conditional distribution as a proposal to update the strain-specific intercept parameters $a_k$ within our MCMC scheme leveraging the Poisson-Gamma conjugacy. 
Define $\phi_k = \exp({a_k})$, and $\mu_{itk} = r_t + s_{t\bmod C} + u_i + x_{itk}\beta_k$. We can then rewrite Equation \ref{eq:multitype-model} as:
\begin{align*}
    y_{itk}|\phi_k,\mu_{itk},e_{it} &\sim \text{Poisson}(e_{it}\phi_k\exp{\left(\mu_{itk})\right)}\\
    \prob(y_{1:I,1:T,k}|\phi_k,\mu_{itk},e_{it}) &= \prod_{i=1}^{I}\prod_{t=1}^{T}\frac{\left(e_{it}\phi_k\exp{(\mu_{itk})}\right)^{y_{itk}}}{y_{itk}!}\exp{\left(-e_{it}\phi_k\exp{(\mu_{itk})}\right)}\\
    &= \left[\prod_{i=1}^{I}\prod_{t=1}^{T}\frac{\left(e_{it}\exp{(\mu_{itk})}\right)^{y_{itk}}}{y_{itk}!}\right]\left[\phi_k^{\sum_{i=1}^{I}\sum_{t=1}^T y_{itk}}\exp{\left(-\phi_k \sum_{i=1}^{I}\sum_{t=1}^T e_{it}\exp{(\mu_{itk})}\right)}\right]\\
    \prob(y_{1:I,1:T,k}|\phi_k,\mu_{itk},e_{it})&\propto \phi_k^{\boldsymbol{Y}_k} \exp{(-\phi_k \boldsymbol{S}_k)}
\end{align*}
where $\boldsymbol{Y}_k = \sum_{i=1}^{I}\sum_{t=1}^T y_{itk}$, and $\boldsymbol{S}_k = \sum_{i=1}^{I}\sum_{t=1}^T e_{it}\exp{(\mu_{itk})}$.

Conjugacy can therefore be achieved by assuming
a prior $\phi_k \sim \text{Gamma}(\alpha^\phi_k,\beta^\phi_k)$ 
which leads to the full conditional distribution:
$$\phi_k|y_{1:I,1:T,k},\bx,\mu_{itk},e_{1:I,1:T} \sim \text{Gamma}(\alpha^\phi_k+\boldsymbol{Y}_k, \beta^\phi_k+\boldsymbol{S}_k)$$

Given that the joint-state sequence $\bx$ are not directly observable from data, we smooth $\boldsymbol{S}_k$ using the marginal posterior probabilities of the hidden states, and marginalize over the state space $\mathcal{X}$ to obtain an unbiased estimate $\boldsymbol{\hat{S}}_k$ for the pseudo-Gibbs sampler. A Metropolis-Hastings accept-reject step with the Jacobian of transformation $\exp{(a_k)}$ is incorporated to guaranty targetting of the correct stationary distribution.

\subsubsection{Gradients and metric tensor for MMALA}
Here, we consider the gradients and metric tensor required for inference. The choice of metric tensor for our MMALA implementation is the observed Fisher information matrix of the log posterior $\boldsymbol{G}(\btheta)=-\boldsymbol{H}(\btheta)$, where $\boldsymbol{H}(\btheta)$ is the Hessian matrix of the log posterior. By the linearity property of differentiation, the gradients and Hessian of the log posterior with respect to $\btheta$ satisfy the following:
\begin{align*}
    \nabla_{\btheta} \ \text{log} \ \prob(\btheta|\by_{1:I,1:T,1:K}) = \nabla_{\btheta} \ \text{log} \ \prob(\by_{1:I,1:T,1:K}|\btheta) +  \nabla_{\btheta} \ \text{log} \ \prob(\btheta).\\
     \nabla_{\btheta}^2 \ \text{log} \ \prob(\btheta|\by_{1:I,1:T,1:K}) = \nabla_{\btheta}^2 \ \text{log} \ \prob(\by_{1:I,1:T,1:K}|\btheta) +  \nabla_{\btheta}^2 \ \text{log} \ \prob(\btheta).
\end{align*}
Since the joint-state sequence $\bx$ has been marginalized out as described in Section \ref{sec:forward filtering}, we now require gradients and Hessian from the log marginal likelihood and the log prior. Exact methods to directly propagate the gradients and Hessian from the log marginal likelihood within the forward algorithm described in Section \ref{sec:forward filtering} have been proposed \citep{qin2000direct, lystig2002exact, turner2008direct}, however, these methods would be computationally intensive in the models being introduced in this paper since model parameters follow the temporal and spatial dimensions of the data being analyzed. In particular, direct propagation of the Hessian within the forward algorithm requires a heavy multi-dimensional array which is practically infeasible in our setting. To obtain a computationally efficient approach that still preserves exact Bayesian posterior inference, we exploit the Fisher and Louis identities.

\subsubsection{The Fisher and Louis identities}
In many statistical models involving latent variables, evaluating the gradients and the Hessian matrix from the complete-data log likelihood is often complicated by the unobserved latent variables. The Fisher and Louis identities provide a theoretical basis for computing the gradients and the Hessian matrix directly from the incomplete-data log likelihood, respectively. For a more detailed discussion and proof of these identities, see \citep{louis1982finding} and Section D.2 of \citep{douc2014nonlinear}. The following Equations \ref{eq:Fishers Identity} and \ref{eq:Louis Identity} are the Fisher and Louis identities, respectively. These identities relate the gradients and Hessian of the incomplete-data log likelihood to the  complete-data log likelihood. 
\begin{align}
     \nabla_{\btheta}\text{log} \ \prob(\by|\btheta) 
     &= \sum_{\bx \in \cX} \left\{\nabla_{\btheta}\text{log} \ \prob(\bx, \by|\btheta)\right\} 
     \prob(\bx|\by,\btheta) =\mathbb{E}_{\bx|\by,\btheta}\left[\nabla_{\btheta}\text{log} \ \prob(\by,\bx|\btheta)\right]. \label{eq:Fishers Identity} \\   
    \nabla^2_{\btheta}\text{log} \ \prob(\by|\btheta) 
    &= \sum_{\bx \in \cX} \left[\nabla^2_{\btheta}\text{log} \ \prob(\bx, \by|\btheta) 
    + \{\nabla_{\btheta}\text{log} \ \prob(\bx, \by|\btheta)\}
      \{\nabla_{\btheta}\text{log} \ \prob(\bx, \by|\btheta)\}^\top \right] 
      \prob(\bx|\by,\btheta) \notag \\
    &\quad - \{\nabla_{\btheta}\text{log} \ \prob(\by|\btheta)\}
      \{\nabla_{\btheta}\text{log} \ \prob(\by|\btheta)\}^\top \notag \\
      &=\mathbb{E}_{\bx|\by,\btheta}\left[\nabla_{\btheta}^2\text{log} \ \prob(\by,\bx|\btheta) + \left\{\nabla_{\btheta}\text{log} \ \prob(\bx, \by|\btheta)\right\}\left\{\nabla_{\btheta}\text{log} \ \prob(\bx, \by|\btheta)\right\}^\top\right] \notag \\
     &\quad - \{\nabla_{\btheta}\text{log} \ \prob(\by|\btheta)\}
      \{\nabla_{\btheta}\text{log} \ \prob(\by|\btheta)\}^\top. \label{eq:Louis Identity}
    % \nabla_{\btheta}\text{log} \ \prob(\by|\btheta) 
    % &= \int \nabla_{\btheta}\text{log} \ \prob(\bx, \by|\btheta) 
    % \prob(\bx|\by,\btheta)\mathrm{d}\bx. \label{eq:Fishers Identity} \\
    % -\nabla^2_{\btheta}\text{log} \ \prob(\by|\btheta) 
    % &=-\int \nabla^2_{\btheta}\text{log} \ \prob(\bx, \by|\btheta) 
    % \prob(\bx|\by,\btheta)\mathrm{d}\bx 
    % + \int \nabla^2_{\btheta}\text{log} \ \prob(\bx|\by, \btheta) 
    % \prob(\bx|\by,\btheta)\mathrm{d}\bx \nonumber \\
    % \nabla^2_{\btheta}\text{log} \ \prob(\by|\btheta) 
    % &= \int \left[\nabla^2_{\btheta}\text{log} \ \prob(\bx, \by|\btheta) 
    % + \{\nabla_{\btheta}\text{log} \ \prob(\bx, \by|\btheta)\}
    %   \{\nabla_{\btheta}\text{log} \ \prob(\bx, \by|\btheta)\}^\top \right] 
    %   \prob(\bx|\by,\btheta)\mathrm{d}\bx \notag \\
    % &\quad - \{\nabla_{\btheta}\text{log} \ \prob(\by|\btheta)\}
    %   \{\nabla_{\btheta}\text{log} \ \prob(\by|\btheta)\}^\top. \label{eq:Louis Identity}
\end{align}

\subsubsection{Decomposing the complete-data log likelihood in the Fisher and Louis identities}
The complete-data log likelihood can be decomposed as:
\begin{align}
    \text{log} \ \prob(\by, \bx|\btheta) &= \text{log} \ \prob(\by|\bx,\btheta) + \text{log} \ \prob(\bx|\btheta), \label{eq:complete-data loglikelihood decomposition}
\end{align}
and inserting this into the Fisher identity implies:
\begin{align}
   \nabla_{\btheta}\text{log} \ \prob(\by|\btheta)&= \mathbb{E}_{\bx|y,\btheta}\left[\nabla_{\btheta}\text{log} \ \prob(\by|\bx,\btheta) + \nabla_{\btheta}\text{log}\prob(\bx|\btheta)\right] \notag \\
 \quad &= \mathbb{E}_{\bx|\by,\btheta}\left[\nabla_{\btheta}\text{log} \ \prob(\by|\bx,\btheta)\right] + \mathbb{E}_{\bx|\by,\btheta} \left[\nabla_{\btheta}\text{log}\prob(\bx|\btheta)\right]. \label{eq:finalFisher decomposition}
\end{align}

Similarly, the Louis identity implies:
\begin{align*}
     \nabla_{\btheta}^2 \ \text{log} \ \prob(\by|\btheta) &= \sum_{\bx \in \cX} \left( \nabla_{\btheta}^2 \ \text{log} \ \prob(\by|\bx,\btheta) + \nabla_{\btheta}^2\text{log} \ \prob(\bx|\btheta)\right)\prob(\bx|y,\btheta) \notag \\
     \quad &+ \sum_{\bx \in \cX} \left(\{\nabla_{\btheta} \ \text{log} \ \prob(\by|\bx,\btheta) + \nabla_{\btheta}\text{log} \ \prob(\bx|\btheta)\}\{\nabla_{\btheta} \ \text{log} \ \prob(\by|\bx,\btheta) + \nabla_{\btheta}\text{log} \ \prob(\bx|\btheta)\}^\top\right)\prob(\bx|\by,\btheta) \notag \\ 
    \quad &- \{\nabla_{\btheta} \ \text{log} \ \prob(\by|\btheta)\}\{\nabla_{\btheta} \ \text{log} \ \prob(\by|\btheta)\}^\top. 
\end{align*}

\subsubsection{Analytic gradients and Hessian of the log likelihood}
Recall the general form of our model in Equation \ref{eq:model-multitype}. The complete-data log likelihood in the context of our model is written as:
\begin{align}
    \text{log} \ \prob(\by_{1:I,1:T,1:K}, \bx|\btheta) &= \sum_{i=1}^{I}\sum_{t=1}^T\sum_{k=1}^K \big\{y_{itk}(a_k + r_t + s_{t\bmod C} + u_i + x_{itk}\beta_k) - \exp(a_k + r_t + s_{t\bmod C} + u_i + x_{itk}\beta_k)e_{it} \notag \\
    &\phantom{=\sum\sum\sum}+ y_{itk} \ \text{log} \ e_{it} - \text{log} \ y_{itk}!\big\} + \text{log} \ \prob(\bx|\bGamma),\label{eq:modelIncomplete loglikelihood} 
\end{align}
where we have used the fact that, in the prior, the joint-state sequence $\bx$ is conditionally independent of all components of $\btheta$ except for $\bGamma$, and so $\prob(\bx|\btheta)=\prob(\bx|\bGamma)$. Now differentiating this complete-data log likelihood with respect to $r_t, s_{t\bmod C}, \text{and} \ u_i$ gives:
\begin{align*}
    \nabla_{r_t} \ \text{log} \ \prob(\by_{1:I,1:T,1:K},\bx|\btheta)=\sum_{i=1}^{I}\sum_{k=1}^K \big\{y_{itk} - \exp(a_k + r_t + s_{t\bmod C} + u_i + x_{itk}\beta_k)e_{it}\big\} \\
    \nabla_{s_{t\bmod C}} \ \text{log} \ \prob(\by_{1:I,1:T,1:K},\bx|\btheta)=\sum_{t:t\bmod C=c}\sum_{i=1}^{I}\sum_{k=1}^K \big\{y_{itk} - \exp(a_k + r_t + s_{t\bmod C} + u_i + x_{itk}\beta_k)e_{it}\big\} \\
        \nabla_{u_i} \ \text{log} \ \prob(\by_{1:I,1:T,1:K},\bx|\btheta)=\sum_{t=1}^T\sum_{k=1}^K \big\{y_{itk} - \exp(a_k + r_t + s_{t\bmod C} + u_i + x_{itk}\beta_k)e_{it}\big\},
\end{align*}
respectively. To apply the Fisher and Louis identities, we define $\eta_{itk}$ as the following posterior expectation, where we smooth $e_{it}\lambda_{itk}$ and marginalize over the state space $\cX$:
\begin{align}
    \eta_{itk}&:=\mathbb{E}_{\bx|y_{1:I,1:T,1:K},\btheta}\left[e_{it}\lambda_{itk}\right] = \sum_{\bx_{it}\in \cX}e_{it}\exp(a_k + r_t + s_{t\bmod C} + u_i + x_{itk}\beta_k) \prob(\bx_{it}|y_{i,1:T,1:K},\btheta) \label{eq:smoothing}, 
\end{align}
where $\prob(\bx_{it}|y_{i,1:T,1:K},\btheta)$ is obtained via the backward algorithm described in Section \ref{sec:outBprob}. 

For a given timepoint $t$, the gradient of the log marginal likelihood with respect to the trend component $r_t$  is given by: 
\begin{align*}
    \nabla_{r_t} \ \text{log} \ \prob(\by_{1:I,1:T,1:K}|\btheta)=\mathbb{E}_{\bx|y_{1:I,1:T,1:K},\btheta}\left[\nabla_{r_t} \ \text{log} \ \prob(\by_{1:I,1:T,1:K},\bx|\btheta)\right]=\sum_{i=1}^{I}\sum_{k=1}^K \big\{y_{itk} - \eta_{itk}\big\}. 
\end{align*}
For season $c$ such that $c=t\bmod C$ for $t=1,2,\dots,T$, the gradient of the log marginal likelihood with respect to the seasonal component $s_{t\bmod C}$  is given by:  
\begin{align*}
      \nabla_{s_{t\bmod C}} \ \text{log} \ \prob(\by_{1:I,1:T,1:K}|\btheta)=\mathbb{E}_{\bx|y_{1:I,1:T,1:K},\btheta}\left[\nabla_{s_{t\bmod C}} \ \text{log} \ \prob(\by_{1:I,1:T,1:K},\bx|\btheta)\right]=\sum_{t:t\bmod C=c}\sum_{i=1}^{I}\sum_{k=1}^K \big\{y_{itk} - \eta_{itk}\big\} 
\end{align*}
For a given spatial location $i$, the gradient of the log marginal likelihood with respect to the spatial component $u_i$  is given by:  
\begin{align*}    
        \nabla_{u_i} \ \text{log} \ \prob(\by_{1:I,1:T,1:K}|\btheta)=\mathbb{E}_{\bx|y_{1:I,1:T,1:K},\btheta}\left[\nabla_{u_i} \ \text{log} \ \prob(\by_{1:I,1:T,1:K},\bx|\btheta)\right]=\sum_{t=1}^T\sum_{k=1}^K \big\{y_{itk} - \eta_{itk}\big\}.
\end{align*}
Given that our chosen metric tensor is the observed Fisher information matrix $G(\btheta)=-H(\btheta)$, we employ the expected curvature component of the Louis identity from the log marginal likelihood, which is a diagonal matrix given our model, to guaranty positive definiteness \citep{girolami2011riemann}, together with the observed Fisher information matrix from the log prior density. This choice is practically and computationally desirable in our model because it ensures that the observed information matrix is sparse and positive definite, which offers fast matrix inversion.
\begin{align*}
    \nabla_{\btheta}^2 \ \text{log} \ \prob(\by|\btheta) &\approx \sum_{\bx \in \cX} \nabla_{\btheta}^2 \ \text{log} \ \prob(\by, \bx|\btheta)\prob(\bx|\by,\btheta)=\mathbb{E}_{\bx|\by,\btheta}\left[\nabla_{\btheta}^2\text{log} \ \prob(\by|\bx,\btheta)\right] + \mathbb{E}_{\bx|\by,\btheta} \left[\nabla_{\btheta}^2\text{log}\prob(\bx|\btheta)\right] \\
               \nabla_{r_t}^2 \ \text{log} \ \prob(\by_{1:I,1:T,1:K}|\btheta)&\approx \mathbb{E}_{\bx|y_{1:I,1:T,1:K},\btheta}\left[\nabla_{r_t}^2 \ \text{log} \ \prob(\by_{1:I,1:T,1:K},\bx|\btheta)\right]=\sum_{i=1}^{I}\sum_{k=1}^K \big\{- \eta_{itk}\big\} \\
    \nabla_{s_{t\bmod C}}^2 \ \text{log} \ \prob(\by_{1:I,1:T,1:K}|\btheta)&\approx\mathbb{E}_{\bx|y_{1:I,1:T,1:K},\btheta}\left[\nabla_{s_{t\bmod C}}^2 \ \text{log} \ \prob(\by_{1:I,1:T,1:K},\bx|\btheta)\right]=\sum_{t:t\bmod C=c}\sum_{i=1}^{I}\sum_{k=1}^K \big\{- \eta_{itk}\big\} \\
        \nabla_{u_i}^2 \ \text{log} \ \prob(\by_{1:I,1:T,1:K}|\btheta)&\approx\mathbb{E}_{\bx|y_{1:I,1:T,1:K},\btheta}\left[\nabla_{u_i}^2 \ \text{log} \ \prob(\by_{1:I,1:T,1:K},\bx|\btheta)\right]=\sum_{t=1}^T\sum_{k=1}^K \big\{- \eta_{itk}\big\}.
\end{align*}
The cross-derivatives in the Hessians of $\text{log} \ \prob(\by_{1:I,1:T,1:K}|\btheta)$ in each of $\br$ and $\bs$ are all zeros since they are independent across time in the likelihood. Similarly for $\bu$ since they are all independent across space in the likelihood. 
The exact analytic gradients and Hessian for the temporal and spatial components from the log priors are provided in Appendix \ref{sec:GradientsAppendix}.

\subsubsection{Sum-to-zero constraints}
The Intrinsic Gaussian Markov Random Fields (IGMRF) used as prior distributions for the trend, seasonal, and spatial components (Equations \ref{eq:trend prior},\ref{eq:seasonal prior}, \ref{eq:spatial prior}, respectively) are improper models and therefore require the sum-to-zero constraint for identifiability. The gradient-based MCMC method presented in this paper adopts the QR-decomposition approach, projecting the temporal and spatial components, their gradients, and their observed Fisher information matrices onto a lower-dimensional unconstrained space using the orthonormal sub-matrix $\bQ_2 \in \mathbb{R}^{n\times n-1}$ obtained from $\bQ$ in the QR-decomposition with $\bQ=[\bQ_1,\bQ_2]$, where $\bQ_1 \in \mathbb{R}^{n\times1}$ is the leftmost column vector in $\bQ$, and the columns of $\bQ_2$ span the null space of the constraint, that is, $\bQ_2^\top \boldsymbol{1}=\boldsymbol{0}$. This projection ensures that the reparameterized parameters live in an unconstrained lower-dimensional space, and that the original parameters satisfy the sum-to-zero constraint while preserving the local geometric structure of the posterior distribution.
Suppose we are interested in sampling $\btheta$ while imposing the sum-to-zero constraint for identifiability, 
\begin{align*}
    \boldsymbol{1}^\top \btheta=0,
\end{align*}
we define the unconstrained subset $\bvartheta$ and reparameterize as follows:
\begin{align*}
    \btheta = \bQ_2\bvartheta,
\end{align*}
where $\bvartheta \in \mathbb{R}^{n-1}$ is unconstrained. Since $\bQ_2$ span the null space of the constraint, $\btheta \in \mathbb{R}^n$ automatically satisfies the sum-to-zero constraint. The unconstrained parameters being sampled are $\bvartheta$, and the projected gradients are:
\begin{align*}
    \nabla_{\bvartheta}\mathcal{L}(\bvartheta) = \bQ_2^\top \nabla_{\btheta}\mathcal{L}(\btheta).
\end{align*}
The projected metric tensor is given by:
\begin{align*}
    \boldsymbol{G}(\bvartheta) = \bQ_2^\top \boldsymbol{G}(\btheta)\bQ_2.
\end{align*}
For a more detailed discussion about sampling on a constraint manifold, see \cite{zappa2018monte}.

\subsection{Bayesian model comparison}
 Bayesian model comparison is typically carried out by computing the Bayes factor \citep{kass1995bayes} which is the ratio of the model evidence $\prob(\by)$ from competing models, defined as:
\begin{align}\label{eq:ModelEvidenceIntegral}
    \prob(\by) = \int_{\boldsymbol{\Theta}} \prob(\by| \boldsymbol{\theta})\prob(\boldsymbol{\theta}) \, \mathrm{d}\btheta.
\end{align}
Here we consider two methods for evaluating the high dimensional integral in Equation \ref{eq:ModelEvidenceIntegral} for our models: bridge sampling and importance sampling. Bridge sampling was originally introduced to directly compute the Bayes factor of two competing models \citep{jeffreys1998theory,kass1995bayes,gronau2017tutorial}. However, this approach would not be efficient here as we have several competing models.
Recent advances show that individual marginal likelihoods can be estimated directly using bridge sampling \citep{overstall2010default}. This involves the construction of a bridge function and the use of samples from both the proposal distribution and the posterior distribution. The importance sampling method, on the other hand, uses samples from the proposal distribution alone. It theoretically provides an unbiased estimate of the marginal likelihood \citep{touloupou2018efficient}, but its performance deteriorates in high dimension due to increasing variance. The bridge sampling approach deals with this variance problem through the use of a well-constructed bridge function \citep{gronau2017tutorial}. 
Table \ref{tab:Summary-of-model-evidence-computation} compares 
the key features of these two model comparison methods, where $\prob(\by|\btheta)$ is the likelihood, $\prob(\btheta)$ is the prior density, $g(\btheta)$ is the proposal distribution, $h(\btheta)$ is the bridge function, $\tilde{\btheta}_i$ is the $i$-th sample from the proposal distribution $g(\btheta)$, and $\btheta^*_j$ is the $j$-th sample from the posterior distribution $\prob(\btheta|\by)$. Note that for the bridge sampling method the bridge function $h(\btheta)$ contains the quantity of interest $\prob(\by)$ and therefore implementation typically requires an iterative scheme. To this end we leverage the implementation from the R package \emph{bridgesampling} \citep{bridgesampling}, where we provide our custom log likelihood, log priors and the necessary parameter constraints. %We refer the reader to \citep{gronau2017tutorial} for a more detailed discussion.

\begin{table}[H]
\centering
        \begin{threeparttable}
    \caption{\textbf{Comparison of methods to estimate marginal likelihood}}
  \begin{tabular}{ c c c l}
        \hline
Method & Estimator & Samples & Bridge function $h(\btheta)$\\
 \hline
Bridge sampling & $\frac{\frac{1}{N_2}\sum_{i=1}^{N_2}\prob(\by|\tilde{\btheta}_i)\prob(\tilde{\btheta}_i)h(\tilde{\btheta}_i)}{\frac{1}{N_1}\sum_{j=1}^{N_1}h(\btheta_j^*)g(\btheta_j^*)}$ & $\Tilde{\btheta}_i \sim g(\btheta)$ & $h(\btheta)\propto\left(\frac{N_1}{N_1+N_2}\prob(\by|\btheta)\prob(\btheta)+\frac{N_2}{N_1+N_2}\prob(\by)g(\btheta)\right)^{-1}$\\
Importance sampling & $\frac{1}{N}\sum_{i=1}^{N}\frac{\prob(\by|\Tilde{\btheta}_i)\prob(\Tilde{\btheta}_i)}{g(\tilde{\btheta}_i)}$ & $\Tilde{\btheta}_i \sim g(\btheta)$ & $h(\btheta) = \frac{1}{g(\btheta)}$ \\
 \hline
   \end{tabular}
   \label{tab:Summary-of-model-evidence-computation}
    \end{threeparttable}
\end{table}

%\begin{align*}
%\hat{p}(y)^{t+1}&=\frac{\frac{1}{n_2}\sum_{i=1}^{n_2}\frac{I_{2,i}}{s_1I_{2,i}+s_2\hat{p}(y)^T}}{\frac{1}{n_1}\sum_{j=1}^{n_1}\frac{1}{s_1I_{1,j}+s_2\hat{p}(y)^T}}\\
%\text{where} I_{1,j}&=\frac{p(y|\theta_j^*)p(\theta_j^*)}{g(\theta_j^*)},\\ I_{2,i}&=\frac{p(y|\Tilde{\theta}_i)p(\Tilde{\theta}_i)}{g(\tilde{\theta}_i)},\\ 
%\theta_j^* &\sim p(\theta|y), \text{and} \ \Tilde{\theta}_i\sim g(\theta).
%\end{align*}

\newpage
\section{Simulation study}
\label{sec:simulationsec}
\begin{figure}[H]
\centerline{\includegraphics[width=1\textwidth]{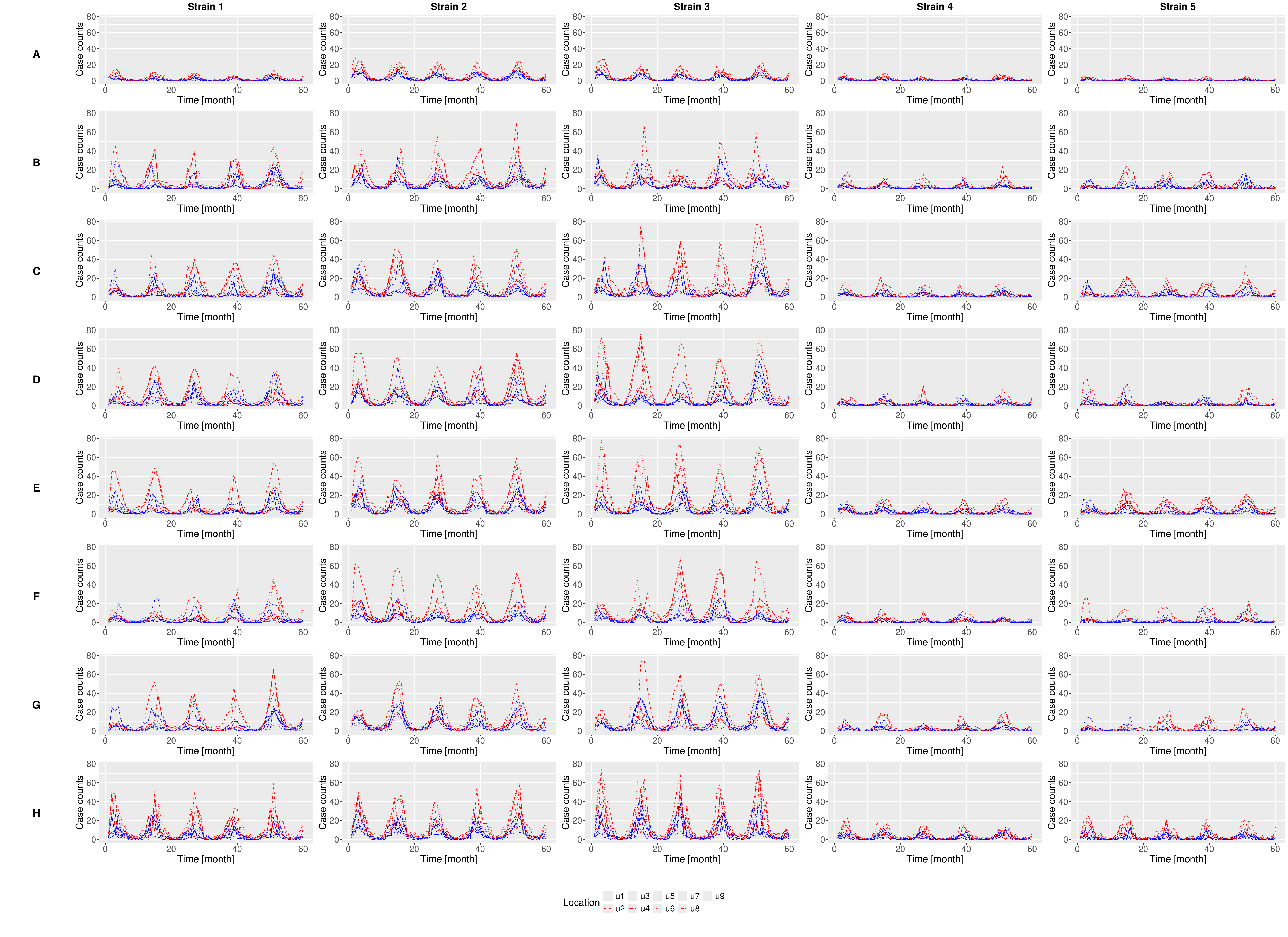}}
\caption{Multi-type model simulations with $K=5$ strains from (A) No epidemic model (B) Independent 1 model (C) Independent 2 model (D) Frank copula 1 model (F) Frank copula 2 model (G) Gaussian factor copula 1 model (H) Gaussian factor copula 2 model, and (H) General-dependent model.}
\label{fig:simulatedData}
\end{figure}

We performed a simulation study in order to assess the validity and scalability of the Bayesian inference methodology described above for each model specified in Table \ref{tab:AllModels}.
We considered that there were five small cities each consisting of approximately 500,000 susceptible individuals on average, and four large cities each consisting of approximately one million susceptible individuals on average. The cities were located in such a way that two cities are considered neighbors if they share a common border. Figure \ref{fig:simulationCities} illustrates the adjacency structure of the hypothetical spatial locations being studied.
Monthly data were simulated with each of the eight models using the joint state-space formulation specified in Equation \ref{eq:multitype-model} for a 5-year period at the nine different geographical locations for $K=5$ strains. The parameters in the spatio-temporal epidemic components and the background per-strain intercepts used for the simulation study were as follows. The simulated intercept parameters for each strain were $a_1=-13.18$, $a_2=-12.31$,
$a_3=-12.49$, $a_4=-13.64$ and $a_5=-13.98$. 
The epidemic effects for each strain were $\beta_1=1.65$, $\beta_2=0.95$,
$\beta_3=1.40$, $\beta_4=1.10$ and $\beta_5=1.70$. 
The Frank copula models used $\psi=6.5$ whereas the Gaussian copula
models used $\rho_{12}=0.80$, $\rho_{13}=-0.85$, $\rho_{14}=0.90$, $\rho_{15}=-0.80$, $\rho_{23}=-0.86$, $\rho_{24}=0.87$, $\rho_{25}=-0.85$, $\rho_{34}=-0.80$, $\rho_{35}=0.80$ and $\rho_{45}=-0.87$.
%listed in Table \ref{tab:multstrainsimulationParams}.
Figure \ref{fig:simulatedData} shows the simulated incidence data
for each strain, location and date, and for each of the eight
models.

%\begin{table}[H]
%\caption{\textbf{Parameters for simulation study}}
%\centering
%\begin{tabular}{lccccc}
%\toprule
%& \multicolumn{5}{c}{Intercepts} \\
%\cmidrule(lr){2-6}
%\textbf{Epidemic status} & $a_1$ & $a_2$ & $a_3$ & $a_4$ & $a_5$ \\
%\midrule
%Absent  & -13.18 & -12.31 & -12.49 & -13.64 & -13.98 \\
%Present & -13.18 & -12.31 & -12.49 & -13.64 & -13.98 \\
%\bottomrule
%\end{tabular}
%\begin{tabular}{lccccc}
%\toprule
%& \multicolumn{5}{c}{Epidemic effects} \\
%\cmidrule(lr){2-6}
%\textbf{Epidemic status} & $\beta_1$ & $\beta_2$ & $\beta_3$ & $\beta_4$ & $\beta_5$ \\
%\midrule
%Absent  & -- & -- & -- & -- & -- \\
%Present & 1.65 & 0.95 & 1.40 & 1.10 & 1.70 \\
%\bottomrule
%\end{tabular}
%\begin{tabular}{lccccccccccc}
%\toprule
%& \multicolumn{11}{c}{Copula parameters} \\
%\cmidrule(lr){2-12}
%\textbf{Copula model}
%& $\rho_{12}$ & $\rho_{13}$ & $\rho_{14}$ & $\rho_{15}$ 
%& $\rho_{23}$ & $\rho_{24}$ & $\rho_{25}$ 
%& $\rho_{34}$ & $\rho_{35}$ & $\rho_{45}$ & $\psi$ \\
%\midrule
%Frank & -- & -- & -- & -- 
%& -- & -- & -- & --
%& -- & -- & 6.5 \\
%Gaussian & 0.80 & -0.85 & 0.90 & -0.80 
%& -0.86 & 0.87 & -0.85 & -0.80
%& 0.80 & -0.87 & -- \\
%\bottomrule
%\end{tabular}
%\label{tab:multstrainsimulationParams}
%\end{table}

For each of these eight simulated datasets, we wanted to perform inference under the same model as was used for simulation.
We tried the two approaches described previously:
on one hand our purpose built approach combining use of MMALA, Gibbs-like and random-walk moves and on the other hand the generic approach implemented by the Stan inference engine.
We found that the former is significantly faster than 
the latter, which can be explained by the fact that 
we provide analytical gradients and observed Fisher information matrices to the samplers in contrast to Stan where numerical gradients via \textit{autodiff} are used. In comparison of computational costs, the total runtime using Stan's HMC No-U-turn sampler with 4 parallel chains on a GPU to run 2000 iterations (1000 warm-up and 1000 sampling, target acceptance statistic $\delta = 0.95$) for the independent model 1 (Equation \ref{eq:Gamma-model1}) was about 48 hours, as compared to 1.09 hours to run 50000 iterations using our purpose-built sampling methods, where major code bottlenecks have been written in \textit{C++}. 
The total run time for the general-dependent model (Equation \ref{eq:Gamma-general}) was approximately 5.9 hours to obtain 50000 MCMC samples with our purpose-built approach, whereas 
Stan would not run in a reasonable amount of time.
The remainder of this article will therefore focus on the
use of our purpose-built approach.

\begin{figure}[H]
\centerline{\includegraphics[width=1\textwidth]{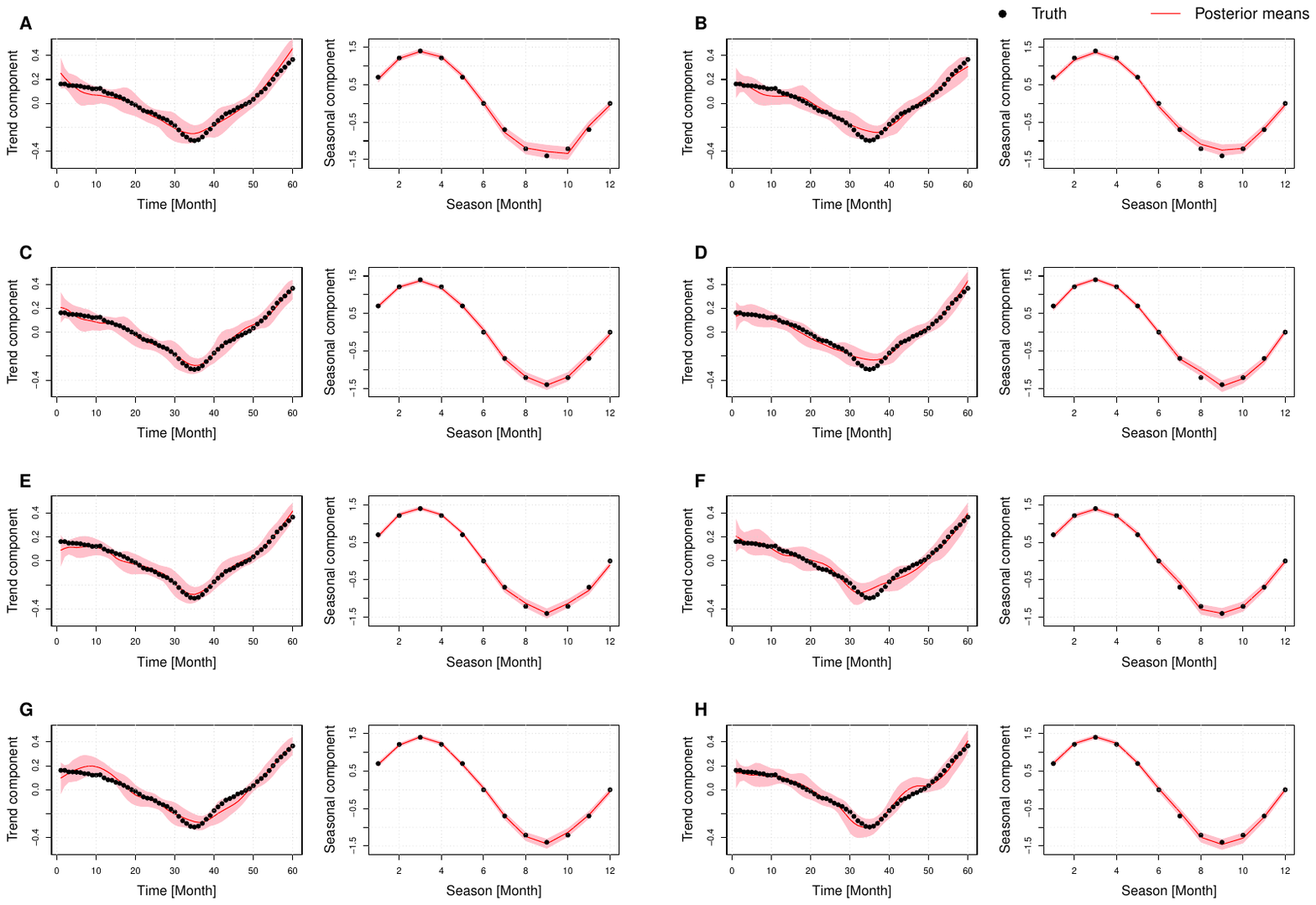}}
\caption{Posterior means, 95\% credible intervals and the true values of the temporal trend and seasonal components from (A) No epidemic model (B) Independent 1 model (C) Independent 2 model (D) Frank copula 1 model (F) Frank copula 2 model (G) Gaussian factor copula 1 model (H) Gaussian factor copula 2 model, and (H) General-dependent model.}
\label{fig:temporalfits}
\end{figure}

\begin{figure}[H]
\centerline{\includegraphics[width=1\textwidth]{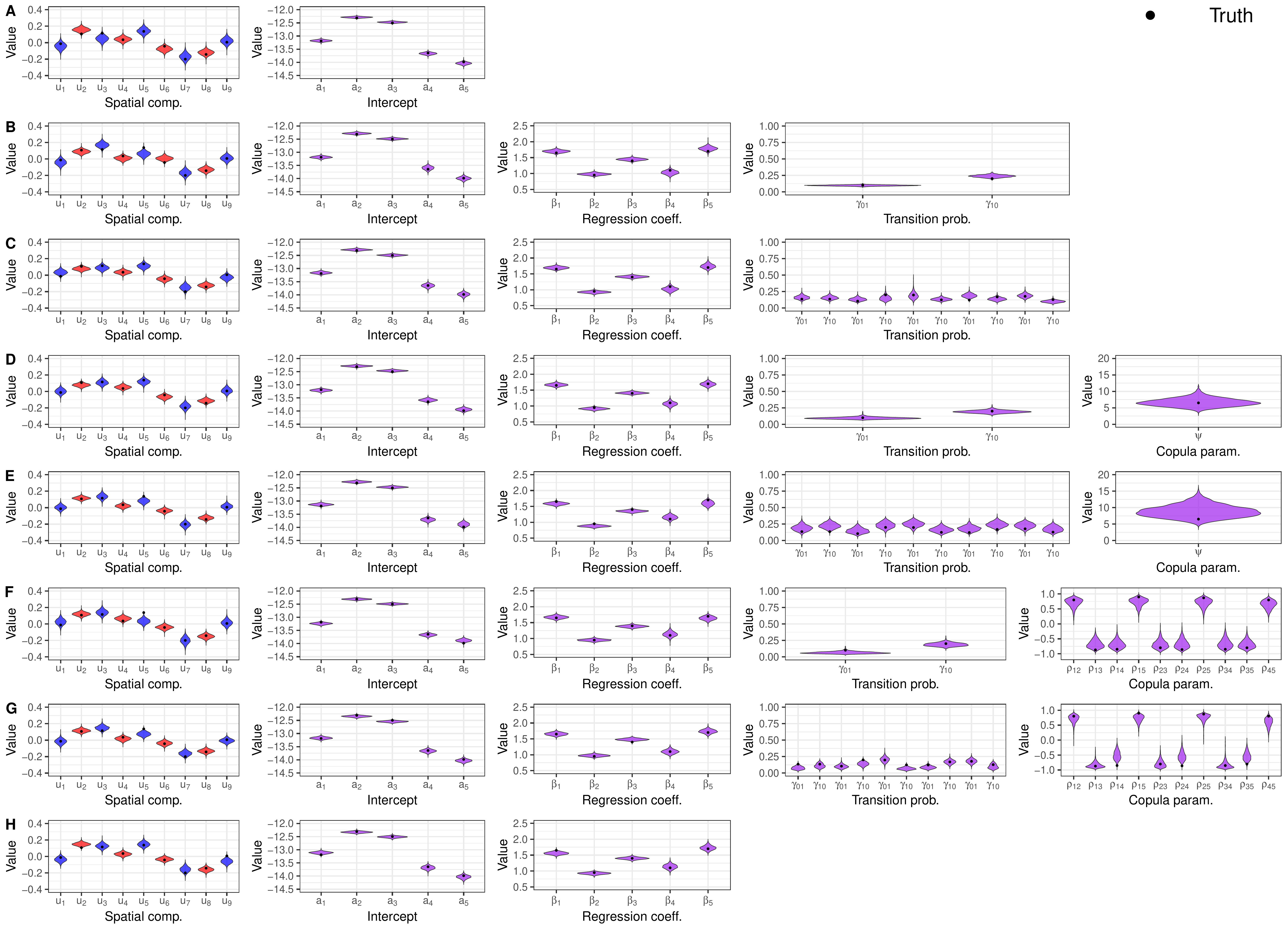}}
\caption{Posterior densities and the true values of the spatial components, intercepts, epidemic effects, transition probabilities and copula parameters from (A) No epidemic model (B) Independent 1 model (C) Independent 2 model (D) Frank copula 1 model (F) Frank copula 2 model (G) Gaussian factor copula 1 model (H) Gaussian factor copula 2 model, and (H) General-dependent model.}
\label{fig:spatialandInterceptfits0To7}
\end{figure}

\begin{figure}[H]
\centerline{\includegraphics[width=1\textwidth]{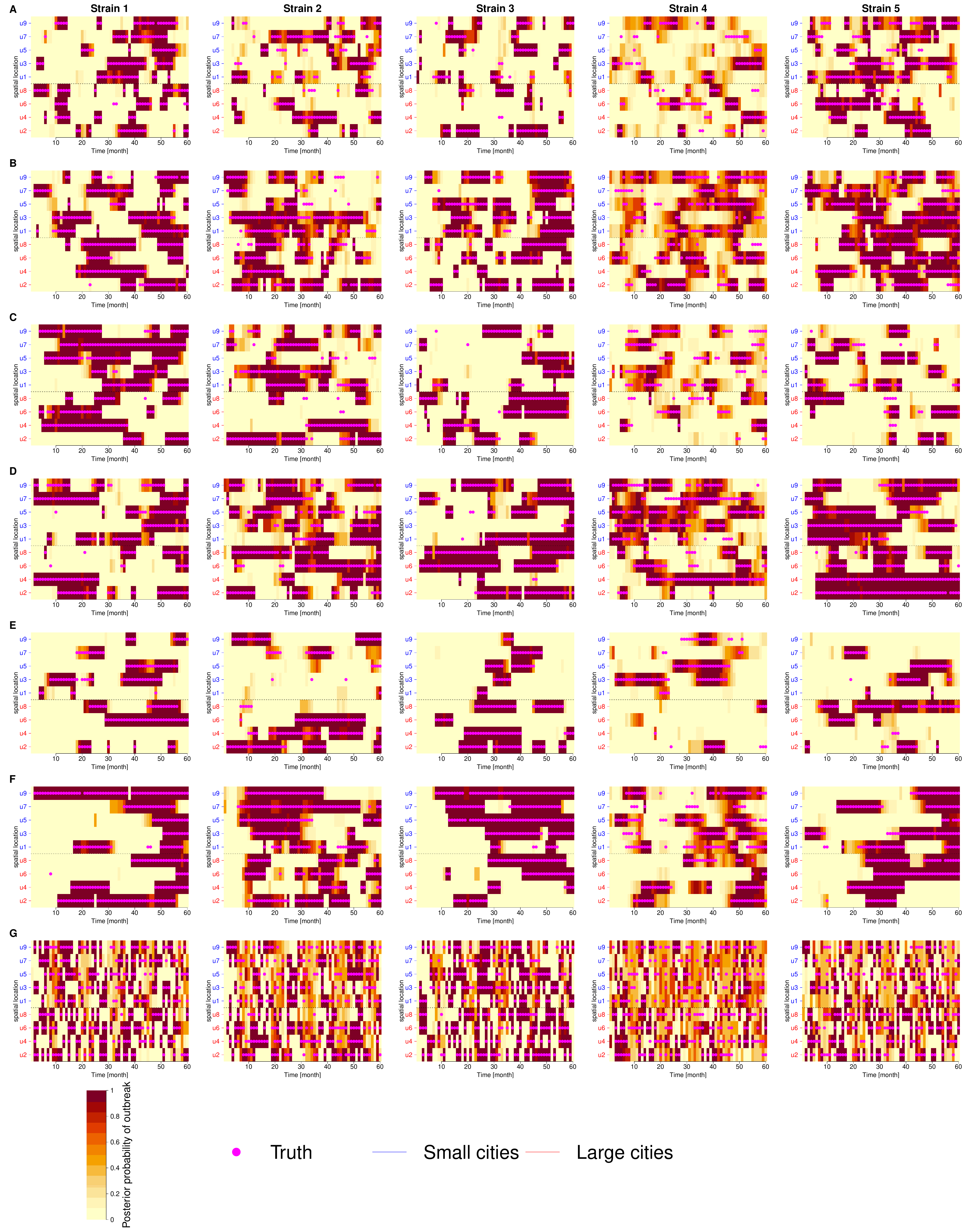}}
\caption{Heatmap showing simulated epidemics per strain and inferred probability of the epidemic state for strains 1 to 5 from (A) Independent 1 model (B) Independent 2 model (C) Frank copula 1 model (D) Frank copula 2 model (E) Gaussian factor copula 1 model (F) Gaussian factor copula 2 model, and (G) General-dependent model.}
\label{fig:alloutbreaks}
\end{figure}

Figure \ref{fig:temporalfits} compares the correct and inferred
values of the temporal trend and seasonal components,
for each of the eight runs. 
Likewise, Figure \ref{fig:spatialandInterceptfits0To7} shows the correct
and inferred values of the spatial components, intercepts,
epidemic effects, transition probabilities and copula
parameters for each of the eight runs.
For all parameters shown in Figures \ref{fig:temporalfits} and \ref{fig:spatialandInterceptfits0To7} 
we find that
the inferred posterior mean is close to the correct value,
and that the 95\% credible interval almost always includes the 
correct value.
Finally, Figure \ref{fig:alloutbreaks} shows the posterior probabilities
of epidemics for each location, date and strain and for each
of the seven models which allow epidemics. These probabilities are
overall in excellent alignment with the correct epidemics that
took place when simulating the data. There are a few examples where
real epidemics have not been identified (false negatives), but this typically happens
for epidemics lasting only one of a few months. 
Such short epidemics would not have had much impact on the
data and therefore are expected to be more difficult to detect.
There are even less instances of inferred epidemics when no real
epidemic had happened (false positives) if we consider 50\% posterior
probability to be the cutoff for calling epidemics, and all of them
had probabilities below 80\%.
The results are less clear for the general-dependent model
(Figure \ref{fig:alloutbreaks}G) due to the more frequent transitions 
between non-epidemic and epidemic states in this model. Nevertheless,
and in spite of the high number of parameters in this model 
(with $K=5$ strains there are 992 parameters in the transition matrix alone),
inference is still satisfactory overall for this model.

%\begin{figure}[!t]
%\centerline{\includegraphics[width=6.5in, height=3.5in, keepaspectratio=true]{Figures/FitUMultMod.pdf}}
%\caption{Posterior densities and the true values of the spatial components, intercepts, epidemic parameters and transition probabilities from Models 0 and 1.}
%\label{fig:spatialandInterceptfits0}
%\end{figure}

\section{Application to multi-type invasive meningococcal disease}
The European Center for Disease Prevention and Control (ECDC) publishes a Surveillance Atlas of Infectious Diseases which contains monthly incidence data on invasive meningococcal disease from 26 European countries \citep{ECDC_Atlas}. The data were available up to December 2023, however, our analysis is restricted to the period from January 2010 to December 2019 to avoid possible confounding factors introduced first by the withdrawal of the United Kingdom (UK) from the European Union and then by the COVID-19 pandemic. Although the dataset includes some missing values, our methodology accommodates this under the assumption of data missing at random as discussed in Section \ref{sec:MissingData}. 
We considered $K=4$ types corresponding to the major serogroups of \textit{Neisseria meningitidis} B, W, Y and C \citep{rouphael2011neisseria}. The ECDC atlas also
includes smaller numbers of cases for other serogroups 
(eg A, X and Z) which were excluded from the analysis.
The raw data is shown in Figure \ref{fig:applicationdata}, from which clear yearly seasonal patterns can be seen. It is also clear that the UK has a high number of disease cases for serogroups B, W and Y, even compared to other countries with a comparably large population such as Germany, France or Italy. This higher incidence may be due to better report rates and surveillance in the UK relative to the other countries, but we are unable to investigate this. 

Annual population counts for each country were available from the Eurostat database \citep{Eurostat}. To derive approximate monthly population counts, we apply linear interpolation to ensure that the population data align with the temporal resolution of the incidence records.
Geographical data on the distances between the capitals of the countries included in our analysis were obtained from the Cshapes R package \citep{gleditsch2010mapping}. Leveraging this data, we classified countries as adjacent if the distance between their capitals was less than or equal to 820 km. This threshold was chosen to ensure that approximately 20\% of pairs of countries were classified as neighbours, providing a balanced representation of geographical proximity in the analysis.

\begin{figure}[H]
\centerline{\includegraphics[width=1\textwidth]{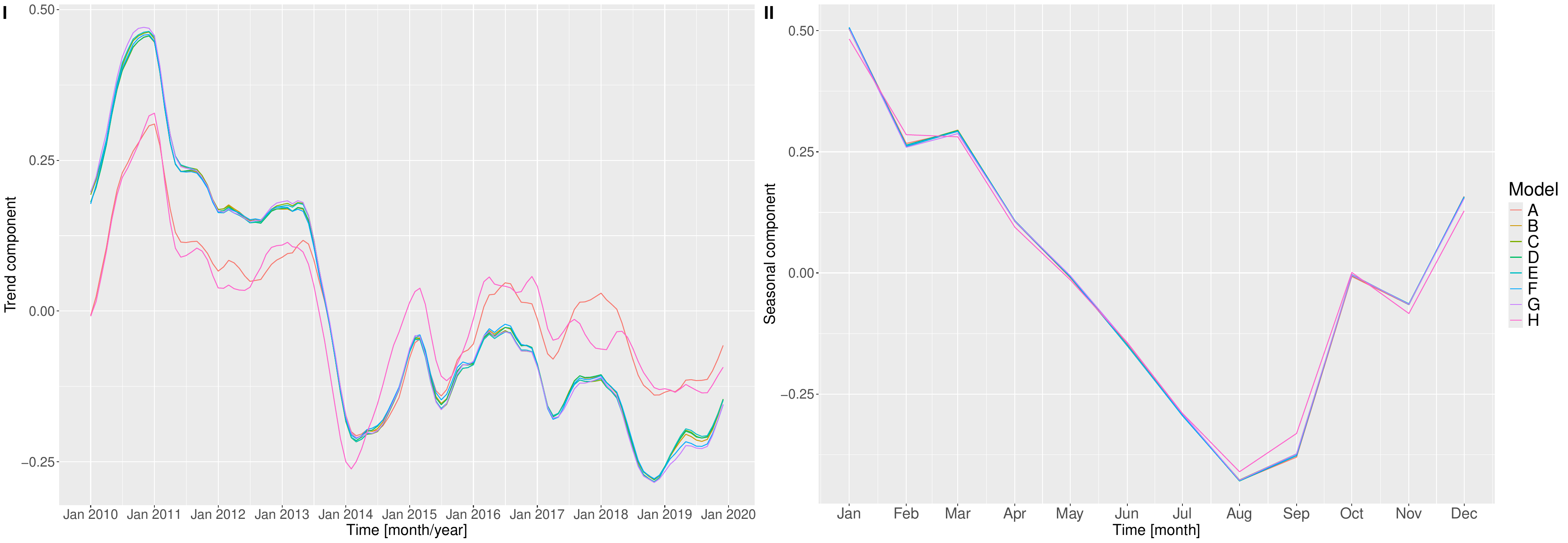}}
\caption{Posterior means for the meningococcal disease application of the trend components, and the seasonal components across all fitted models. (A) No epidemic model (B) Independent 1 model (C) Independent 2 model (D) Frank copula 1 model (F) Frank copula 2 model (G) Gaussian factor copula 1 model (H) Gaussian factor copula 2 model, and (H) General-dependent model.}
\label{fig:applicationtrendseasonalcomps}
\end{figure}

\begin{figure}[H]
\centerline{\includegraphics[width=1\textwidth]{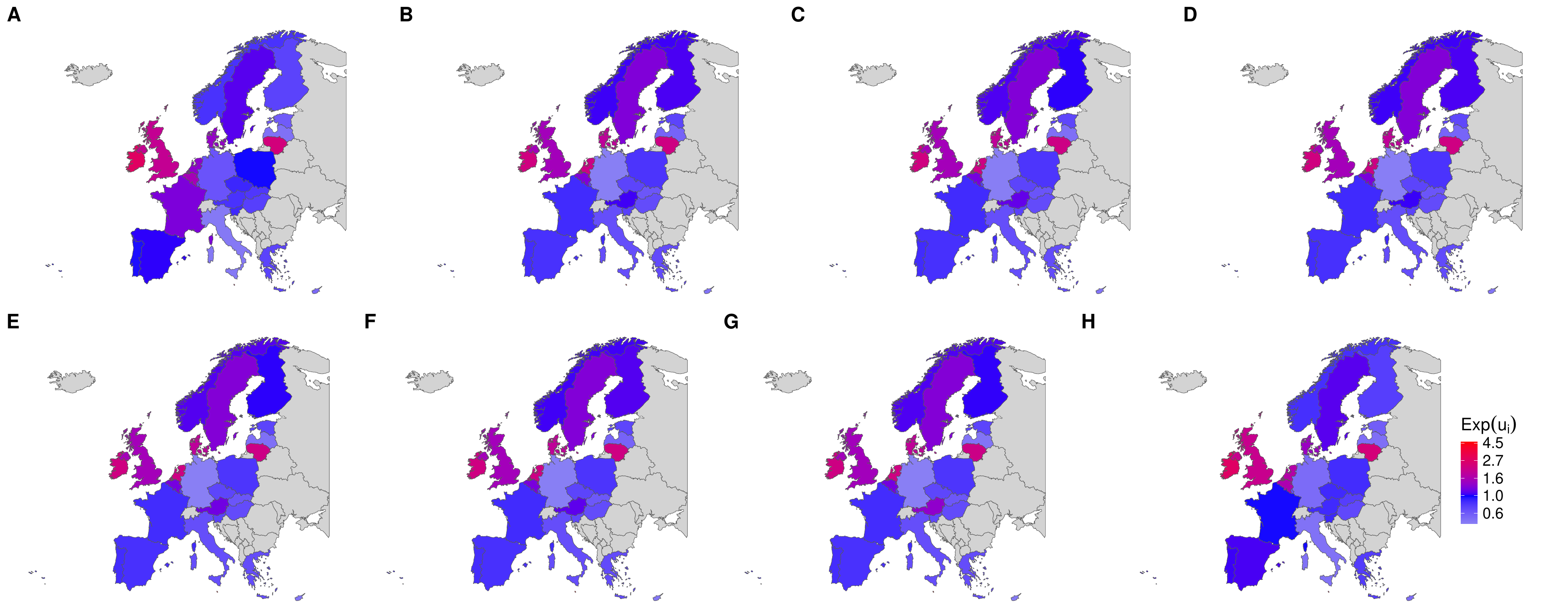}}
\caption{Posterior median relative risks (compared to the geometric mean risk, the mean of the log of the risk) for the meningococcal disease application from (A) No epidemic model (B) Independent 1 model (C) Independent 2 model (D) Frank copula 1 model (F) Frank copula 2 model (G) Gaussian factor copula 1 model (H) Gaussian factor copula 2 model, and (H) General-dependent model.}
\label{fig:applicationrelativeriskmap}
\end{figure}

We performed Bayesian inference on this data separately under
each of the eight models listed in Table \ref{tab:AllModels}.
We run two MCMC chains for all fitted models, with 200000 iterations per chain. MCMC convergence and diagnostics were assessed using the coda R package \citep{plummer2006coda} and found to be satisfactory.
Figure \ref{fig:applicationtrendseasonalcomps}
shows the temporal trend and seasonal component inferred for each model. The seasonal component is almost identical across all eight models. The temporal component is almost the same for the first six epidemic models, but differs evidently for the model without epidemics and the general-dependent model, with lower values in 2011-2013 and 
higher values in 2016-2019. 
Figure \ref{fig:applicationrelativeriskmap} shows the
spatial component of the disease incidence risk, which looks
very similar across the seven models with epidemics but
different in the model without epidemics, especially
for France, the Netherlands, Denmark, Norway and Sweden. 
These differences for the temporal and spatial components
of the model without epidemics can be explained by the fact that under this model any increase in incidence caused by epidemics can only be explained by an increase in other model components. In the models with epidemics, the temporal,
seasonal and spatial components look similar to a 
previous analysis of the untyped data
\citep{ADEOYE2026100879}, as would be expected given that these components affect all strains equally.

\begin{figure}[H]
\centerline{\includegraphics[width=\textwidth, height=\textheight, keepaspectratio]{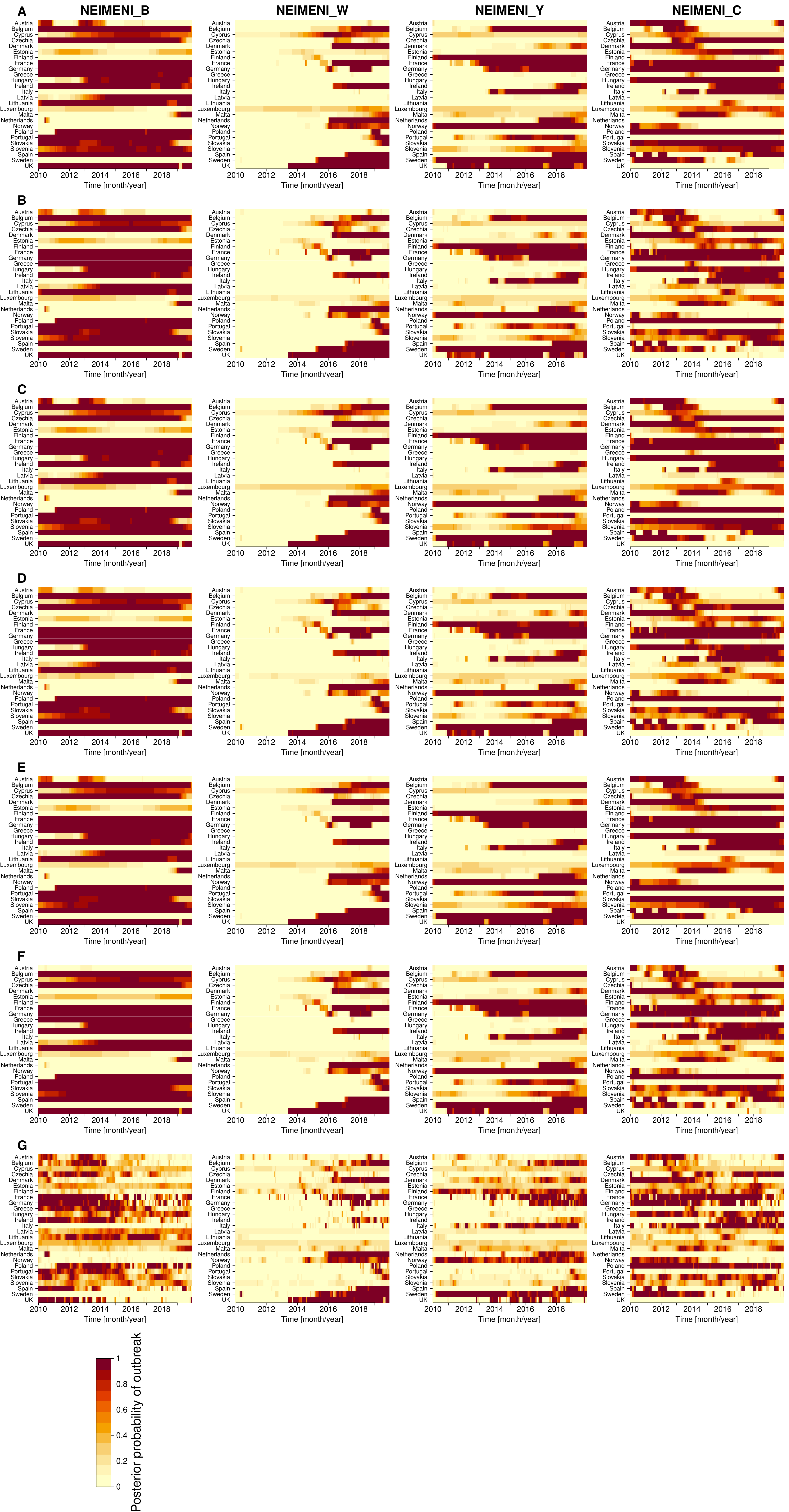}}
\caption{Heat-maps showing for the meningococcal disease application the posterior probability of the epidemic state using (A) Independent 1 model (B) Independent 2 model (C) Frank copula 1 model (D) Frank copula 2 model (E) Gaussian factor copula 1 model (F) Gaussian factor copula 2 model, and (G) General-dependent model.}
\label{fig:applicationheatmaps}
\end{figure}

Figure \ref{fig:applicationheatmaps} shows the inferred
probabilities of the epidemic state in each country over time
for each of the four types and for each of the seven epidemic models, with the notable exception of the general-dependent model which will be discussed later. These results are relatively
consistent across models and show very clear dynamic
differences between the four serogroups which could not have been investigated in an analysis
of untyped data \citep{ADEOYE2026100879}.
For the serogroup B countries typically stay in the endemic or epidemic state over many years, often spanning
the whole study period. Serogroup W was endemic
across all countries at the start of the study period in 2011,
with the visually apparent exception of 
Cyprus and Luxembourg being caused by low 
populations and case numbers in these countries resulting
in high uncertainty for all types and dates. 
Serogroup W switched to the epidemic state in several countries
between 2013 and 2017, including Denmark, France, Ireland, 
Spain, Sweden and the UK, and for most of these countries
stayed in the epidemic state until the end of 2019.
Serogroups Y and C show a more mixed picture, with
some countries such as France or Germany being in the epidemic state almost throughout and other countries switching
between the two states several times during the study
period, without much synchrony between countries. 
These results are in good agreement with a previous 
descriptive analysis of the evolution of invasive 
meningococcal disease in Europe \citep{nuttens2022evolution}. 
In particular, this study noted the rise in serogroup W
across several European countries
with an overall increase of 517\% from 2008 to 2017.
Our analysis found that the UK was the first country
to be affected in 2013, as previously noted \citep{ladhani2015increase,krone2019increase}.
For serogroup Y, several countries (Belgium, Denmark, Finland,
France, the Netherlands, Norway, Sweden and the UK)
were reported to have notably higher incidences than other countries in 
2017 \citep{nuttens2022evolution} and our results show
that these countries are in the epidemic state at that time.

Figure \ref{fig:applicationposteriorpredictives} shows the 95\% posterior predictive credible intervals \citep{gelman1996posterior} for the number of cases for each strain. The real data are almost always contained in the 95\% posterior predictive credible intervals across all models except the model without epidemics, particularly for serogroup W. 
Table \ref{tab:applicationEvidence} contains the 
log marginal likelihood for each model estimated by both
the importance sampling \citep{touloupou2018efficient} and bridge sampling 
\citep{gronau2017tutorial} methods based on 
50 repeated evaluations in an attempt to quantify the Monte Carlo standard error of our estimate.
The two methods are in good agreement, both finding
that model Frank copula 1 is most likely, with
only the two Independent 1 and Gaussian copula 1 have 
non-negligible posterior probability. However, the importance
sampling method gives a more balanced probability between
these three models compared to the bridge sampling method.
When the correlation parameters of the Frank copula 1 
and Gaussian copula 1 models are small, they reduce
to the Independent 1 model which explains why it is
difficult here to distinguish between them. The three
models with separate strain marginal distributions have
lower posterior probabilities, indicating that there
is no evidence against all strains having the same
marginal behaviour. Finally, the general-dependent model
had the lowest log marginal likelihood, indicating
that the different results from this model 
(Figures \ref{fig:applicationtrendseasonalcomps}
and \ref{fig:applicationheatmaps}) should not be trusted.
This model can theoretically reduce to the same
as the best models, but is handicapped by the 
large number of parameters in its transition matrix
($2^K(2^K-1)=240$ for $K=4$).

\begin{table}[H]
\centering
        \begin{threeparttable}
    \caption{\textbf{Model comparison for the multi-type meningococcal disease application with mean and standard deviations reported based on 50 repeated evaluations of the log marginal likelihood.}}
  \begin{tabular}{c cc cc}
        \hline
    & \multicolumn{2}{c}{Log marginal likelihood} & \multicolumn{2}{c}{Posterior model probability} \\
\cline{2-3} \cline{4-5}
Model type & IS (SD) & BS (SD) & IS (SD) & BS (SD) \\
 \hline
No epidemic&-17976.50 (1.03)&-18014.07 (0.17)& 0.0000 (0.00)& 0.0000 (0.00) \\
Independent 1& -15071.63 (0.96) &-15109.78 (0.16) & 0.3079 (0.22)& 0.0253 (0.00) \\
Independent 2&-15125.94 (1.48)&-15164.38 (0.19) &0.0000 (0.00)& 0.0000 (0.00) \\
Frank copula 1&\textbf{-15071.46} (1.20)& \textbf{-15106.12} (0.23)&\textbf{0.3911} (0.28)&\textbf{0.9484} (0.01) \\
Frank copula 2&-15126.72 (1.50)& -15162.31 (0.18)&0.0000 (0.00) &0.0000 (0.00) \\
Gaussian factor copula 1&-15071.76 (1.39)&-15109.75 (0.20)&0.3009 (0.26) &0.0262 (0.00) \\
Gaussian factor copula 2&-15123.19 (1.81)&-15162.68 (0.25)&0.0000 (0.00)&0.0000 (0.00) \\
General-dependent&-18266.30 (22.29)&-15667.63 (2.15)&0.0000 (0.00)&0.0000 (0.00) \\
 \hline
   \end{tabular}
   \begin{tablenotes}
   \item IS: importance sampling, BS: bridge sampling, SD: standard deviation.
   \end{tablenotes}
   \label{tab:applicationEvidence}
    \end{threeparttable}
\end{table}

\section{Discussion}

We have presented a new framework for the analysis of spatio-temporal epidemic data,
which builds upon previous work \citep{knorr2003hierarchical,spencer2011detection,ADEOYE2026100879} 
but has the key new feature of being able to handle multi-type data. We designed
several models in which multiple types can cause epidemics with different dependency
structures, exploiting probability theory on copulas \citep{nelsen2006introduction}.
For each model we showed how Bayesian inference can be performed, and also how
Bayesian model comparison can reveal the most appropriate model for analysis.

As with any model design, the basic formulation in Equation \ref{eq:multitype-model}   makes a number
of assumptions that are worth discussing. 
In contrast to previous work \citep{ADEOYE2026100879}, the models presented in this study incorporate strain-specific intercepts ($a_k$), which separates the intercept from the trend. This separation allows the intercepts to vary across strains while keeping the temporal components ($r_t$ and $s_t$) fixed across strains. 
However, all strains share the same temporal components. Allowing each strain to have
a separate temporal trend component ($r_{tk}$) would in principle be straightfoward and
potentially allow the estimation of interesting differences between strain dynamics. However, it would 
increase the number of parameters to estimate and they might suffer from identifiability
issues vis-a-vis the epidemic terms ($x_{itk}$). We therefore did not attempt this extended
parametrisation, although there might be some settings where it would be useful.
Another potentially interesting avenue for future work would be to use a similar formulation
to analyse multi-host data. This would be useful to study pathogens
that can infect several species, for example \textit{Salmonella enterica} or 
\textit{Campylobacter jejuni}, but also to integrate relevant data on the host,
for example the age of individuals infected with \textit{Neisseria meningitidis} which
is known to be associated with infection numbers in datasets such as the one we 
analysed here \citep{nuttens2022evolution}.

For the Bayesian inference we initially attempted to use 
the probabilistic programming software Stan \citep{carpenter2017stan}. 
It provides a versatile gradient-based 
MCMC sampler known as the No-U-Turn sampler (NUTS) which has contributed to the widespread accessibility of Bayesian computation to researchers from various disciplines. 
However, we found Stan to be inefficient with the models we proposed due to high dimensionality and the burden of computing the entire gradients of the log posterior in every leapfrog step via automatic differentiation. We therefore developed a bespoke MCMC sampling scheme using analytically derived gradients and observed Fisher information matrices to support efficient posterior exploration using a MMALA scheme \citep{girolami2011riemann}. This allowed us to analyse
datasets at a scale and speed that would not have been possible otherwise. 

To calculate the model evidences, we used both importance sampling \citep{touloupou2018efficient}
and bridge sampling \citep{gronau2017tutorial}. Whilst the two methods were in agreement as to the best fitting model, we see a substantial disagreement in the overall level of the log marginal likelihood between the two methods that was much larger than the Monte Carlo standard errors. For the general-dependent model, the failure of the importance sampling method to produce comparable estimates probably stems from high-dimensionality of the posterior (for example, it has 240 transition parameters versus only 12 transition parameters in the Gaussian factor copula 2 model), and the fact that sampling the transmission probability vectors from a heavy-tail multivariate \textit{t} distribution does not constrain them to lie within the 16 dimensional simplex, even when the mean vector satisfies this constraint. We attempted to solve the challenge by imposing a constraint to enforce the transition matrix to be a stochastic matrix, however, this only slightly improved the estimate and its associated Monte Carlo standard error. In contrast, the bridge sampling scheme performed well because it uses samples from both the posterior distribution and the proposal distribution within a well constructed bridge function whilst taking into account the required constraints of all model parameters, and therefore produces more reliable estimates.

In conclusion, the work presented here contains several innovations in the
modelling of dependency between types of an infectious disease, in the Bayesian inference
of parameters using Monte-Carlo techniques and in the model comparison methodology.
Beyond our own goals, these methodological developments are likely to be useful to other scientific
researchers developing methods to analyse large multi-factorial infectious disease datasets under 
complex parametric models. Furthermore, our inferential framework is implemented into an open 
source software which is generally applicable to a wide range of spatio-temporal epidemic data.

\clearpage
\appendix
\section{Appendix}
\setcounter{equation}{0}\renewcommand\theequation{A\arabic{equation}}
\subsection{Temporal trend component} \label{sec:trendAppendix}
To estimate a smooth background temporal trend, the trend component, $r_t$, is assumed to follow a second-order Gaussian random walk prior given by 
\begin{equation*} 
r_t | r_{t-1}, r_{t-2} = 2r_{t-1} - r_{t-2} + \epsilon_t, 
\end{equation*}
for $t = 3, ..., T$, where $\epsilon_t \sim \mathcal{N}(0, \kappa_r^{-1})$. The components of the temporal trend are forced to sum to zero, in contrast to the formulation in \citep{ADEOYE2026100879} where they also act as an intercept in the model.

The joint prior density of the trend components $\boldsymbol{r}$ is written as
\[
    \prob(\boldsymbol{r}|\kappa_r) \propto \kappa_r^{\frac{T-2}{2}} \exp\left(-\frac{\kappa_r}{2}\sum_{t = 3}^{T}(r_t - 2r_{t-1} + r_{t-2})^2\right),
\]
where $\kappa_r$ is the precision parameter, with the constraint that $\sum_{t=1}^T=0$. An alternative formulation of this density is written as
\begin{equation} \label{eq:trend prior}
\prob(\boldsymbol{r}|\kappa_r) \propto \kappa_r^{\frac{T-2}{2}} \exp\left(-\frac{\kappa_r}{2}\boldsymbol{r}^\top\boldsymbol{R_r}\boldsymbol{r}\right)\mathbbm{1}_{ (\boldsymbol{r}^\top\boldsymbol{1}=0)}
\end{equation}
where $\boldsymbol{R_r}$ is a $T\times T$ structure matrix of the second-order Gaussian random walk (RW2) prior, describing the neighborhood structure of the components:

\[
\boldsymbol{R_r}=\begin{bmatrix*}[r]
1 & -2 & 1 & 0 & 0 & 0 & \ldots & 0 & 0 & 0\\
-2 & 5 & -4 & 1 & 0 & 0 & \ldots & 0 & 0 & 0\\
1 & -4 & 6 & -4 & 1 & 0 & \ldots & 0 & 0 & 0\\
0 & 1 & -4 & 6 & -4 & 1 & \ldots & 0 & 0 & 0\\
\vdots & \vdots & \vdots & \vdots & \vdots & \vdots & \ddots & \vdots & \vdots & \vdots\\
0 & 0 & 0 & 0 & 0 & 0 & \ldots & 1 & -2 & 1\\
\end{bmatrix*}
\]

\subsection{Seasonal component} 
\label{sec:seasonalAppendix}
We propose a first-order cyclic Gaussian random walk model for the seasonal components, $s_t$. This model is formulated such that the seasonal components are repeated throughout the time period, enabling the seasonal component to capture a static repeating pattern. The number of components in the cycle is chosen based on the structure of the available dataset and the epidemiology of the disease under study. 
\[\boldsymbol{s} = (s_1, s_2, \dots , s_{C}),\]
\[s_c - s_{c-1} \sim \mathcal{N}(0, \kappa_{s}^{-1}),\]
$c = 2, \dots , C$ and $s_1 - s_{C} \sim \mathcal{N}(0, \kappa_{s}^{-1})$, so that $s_1$ and $s_{C}$ are considered neighbouring components. The joint density for this prior is written as 
\begin{equation} \label{eq:seasonal prior}
    \prob(\boldsymbol{s}|\kappa_s) \propto ({\kappa_s})^{\frac{C - 1}{2}}\exp\left(-\frac{\kappa_s}{2}\boldsymbol{s}^\top\boldsymbol{R_c}\boldsymbol{s}\right)\mathbbm{1}_{ (\boldsymbol{s}^\top\boldsymbol{1}=0)}
\end{equation}
where $C$ represents the number of components in the cycle, and $\boldsymbol{R_c}$ represents a $C \times C$ structure matrix for the cyclic first-order random walk model (cRW1). The resulting structure matrix is given by: 

\begin{equation*} \label{eq:R_c}
\boldsymbol{R_c}=\begin{bmatrix*}[r]
2 & -1 & 0 & 0 & 0 & 0 &\ldots & -1\\
-1 & 2 & -1 & 0 & 0 & 0 & \ldots & 0\\
0 & -1 & 2 & -1 & 0 & 0 &\ldots & 0\\
0 & 0 & -1 & 2 & -1 & 0 & \ldots & 0\\
\vdots & \vdots & \vdots & \vdots & \vdots & \ddots & \vdots &\vdots\\
-1 & 0 & 0 & 0 & 0 & \ldots & -1 & 2\\
\end{bmatrix*}
\end{equation*}
The sum-to-zero constraint, $\sum_{c=1}^{C} s_c = 0$, is imposed for identifiability of the seasonal components. We implement $C=12$ for both the simulation and application. Thus, each component representing a month in a year and the cycle is forced to repeat over the entire period of the dataset.

\subsection{Spatial component} \label{sec:spatialAppendix}
For the spatial components, $u_i$, we assume a Gaussian intrinsic autoregression, also known as an intrinsic Gaussian Markov Random Field (IGMRF). This model is commonly used in disease mapping in the situation of low counts \citep{besag1991bayesian}. 
\[u_{i}|u_{-i} \sim \mathcal{N}\left(\sum_{j \in n(i)}\frac{u_{j}}{|n(i)|}, \frac{\sigma^2}{|n(i)|}\right),\]
where $n(i)$ is the set of indices of locations that neighbour location $i\in\{1,\dots,I\}$. The joint density of the spatial components $\boldsymbol{u}$ is written as
\begin{equation} \label{eq:spatial prior}
    \prob(\boldsymbol{u}|\kappa_u) \propto \left({\frac{\kappa_u}{2\pi}}\right)^{\frac{I-k}{2}}\exp\left(-\frac{\kappa_u}{2}\boldsymbol{u}^\top\boldsymbol{R_u}\boldsymbol{u}\right)\mathbbm{1}_{ (\boldsymbol{u}^\top\boldsymbol{1}=0)}
\end{equation}
where $\boldsymbol{R_u}$ is the structure matrix derived from a given adjacency matrix describing the connectivity of the geographical locations under study. $R_u$ has rank $I-k$, with $k$ being its rank deficiency. The elements of $\boldsymbol{R}$ are
\[
R_{ij} =  \begin{cases}
|n(i)| &\text{if} \ i = j, \\
-1 &\text{if} \ j \in n(i), \\
0 &\text{otherwise}. \end{cases}
\]

Let $\boldsymbol{Q}$ denote the precision matrix for an intrinsic Gaussian Markov random field, so that $\boldsymbol{Q}^{-1} = \boldsymbol{\Sigma}$ is its variance-covariance matrix. The matrix $\boldsymbol{Q}$ is rank deficient due to the sum to zero constraint of the structure matrix, $\boldsymbol{R_u}$. This implies the non-invertibility of $\boldsymbol{Q}$, thus, $\boldsymbol{\Sigma}$ is undefined. Since IGMRFs are improper distributions, they cannot be generative models for data but can be used as priors for modelling \citep{rue2005Gaussian}. Here, we derive the elements of the precision matrix for a first-order IGMRF for spatial components.
\[
\text{Prec}(u_{i})=Q_{ii}=\kappa_u|n(i)|%\frac{|n(i)|}{\sigma^2}
\]

Assuming all spatial components have mean 0,
\[\mathbb{E}(u_{i}|\boldsymbol{u}_{-i})= \sum_{j \in n(i)}\frac{u_{j}}{|n(i)|}= -\frac{1}{Q_{ii}}\sum_{j \in n(i)}Q_{ij}u_{j}\]

 \[\Rightarrow Q_{ij}=  
\begin{cases}
    -\kappa_u & j\in n(i)\\
    \kappa_u |n(i)| & j=i\\
    0 & \text{otherwise.}
\end{cases}\]
Hence, $\boldsymbol{Q} = \kappa_u \times \boldsymbol{R_u}$.

\subsection{Copula in the case with two strains}
\label{sec:copulaAppendixK2}
For $K=2$, define $u$ and $v$ as the marginal probabilities that strains 1 and 2 transit to state 1 in the next step, respectively. The four joint probabilities are:

\begin{align*}
    \prob(0,0)&=1-u-v+\cC(u,v)\\
    \prob(0,1)&=v-\cC(u,v)\\
    \prob(1,0)&=u-\cC(u,v)\\
    \prob(1,1)&=\cC(u,v)
\end{align*}
The elements of the joint transition matrix are therefore:

\begin{align*}
\text{From state} \ (0,0), \ u=p, v=p \\
    (0,0)\rightarrow(0,0) &= 1-p-p+\cC(p,p)\\
    (0,0)\rightarrow(0,1) &= p-\cC(p,p)\\
    (0,0)\rightarrow(1,0) &= p-\cC(p,p)\\
    (0,0)\rightarrow(1,1) &= \cC(p,p) \\
\text{From state} \ (0,1), \ u=p, v=1-q \\
    (0,1)\rightarrow(0,0) &= 1-p-(1-q)+\cC(p,1-q)\\
    (0,1)\rightarrow(0,1) &= 1-q-\cC(p,1-q)\\
    (0,1)\rightarrow(1,0) &= p-\cC(p,1-q)\\
    (0,1)\rightarrow(1,1) &= \cC(p,1-q) \\ 
\text{From state} \ (1,0), \ u=1-q, v=p \\
    (1,0)\rightarrow(0,0) &= 1-(1-q)-p+\cC(1-q,p)\\
    (1,0)\rightarrow(0,1) &= p-\cC(1-q,p)\\
    (1,0)\rightarrow(1,0) &= 1-q-\cC(1-q,p)\\
    (1,0)\rightarrow(1,1) &= \cC(1-q,p) \\
\text{From state} \ (1,1), \ u=1-q, v=1-q \\
    (1,1)\rightarrow(0,0) &= 1-(1-q)-(1-q)+\cC(1-q,1-q)\\
    (1,1)\rightarrow(0,1) &= 1-q-\cC(1-q,1-q)\\
    (1,1)\rightarrow(1,0) &= 1-q-\cC(1-q,1-q)\\
    (1,1)\rightarrow(1,1) &= \cC(1-q,1-q) 
\end{align*}

\subsection{Copula in the case with three strains}
\label{sec:copulaAppendixK3}
For $K=3$, define $u$, $v$ and $w$ as the marginal probabilities that strains 1, 2 and 3 transit to state 1 in the next step, respectively. The eight joint probabilities are given by the standard inclusion-exclusion representation of the joint cumulative distribution function of a 3-dimensional copula:
\begin{align*}
    \prob(0,0,0)&=1-u-v-w+\cC(u,v,1)+\cC(u,1,w)+\cC(1,v,w)-\cC(u,v,w)\\
    \prob(0,0,1)&=w-\cC(u,1,w)-\cC(1,v,w)+\cC(u,v,w)\\
    \prob(0,1,0)&=v-\cC(u,v,1)-\cC(1,v,w)+\cC(u,v,w)\\
    \prob(1,0,0)&=u-\cC(u,v,1)-\cC(u,1,w)+\cC(u,v,w)\\
    \prob(0,1,1)&=\cC(1,v,w)-\cC(u,v,w)\\
    \prob(1,0,1)&=\cC(u,1,w)-\cC(u,v,w)\\
    \prob(1,1,0)&=\cC(u,v,1)-\cC(u,v,w)\\
    \prob(1,1,1)&=\cC(u,v,w)
\end{align*}
The elements of the joint transition matrix are:
\begin{align*}
\text{From state} \ (0,0,0), \ &u=p, v=p, w=p \\
(0,0,0)\rightarrow(0,0,0) &= 1-p-p-p+\cC(p,p,1)+\cC(p,1,p)+\cC(1,p,p)-\cC(p,p,p)\\
(0,0,0)\rightarrow(0,0,1) &= p-\cC(p,1,p)-\cC(1,p,p)+\cC(p,p,p)\\
(0,0,0)\rightarrow(0,1,0) &= p-\cC(p,p,1)-\cC(1,p,p)+\cC(p,p,p)\\
(0,0,0)\rightarrow(1,0,0) &= p-\cC(p,p,1)-\cC(p,1,p)+\cC(p,p,p)\\
(0,0,0)\rightarrow(0,1,1) &= \cC(1,p,p)-\cC(p,p,p) \\
(0,0,0)\rightarrow(1,0,1) &= \cC(p,1,p)-\cC(p,p,p) \\
(0,0,0)\rightarrow(1,1,0) &= \cC(p,p,1)-\cC(p,p,p) \\
(0,0,0)\rightarrow(1,1,1) &= \cC(p,p,p) \\
\text{From state} \ (0,0,1), \ &u=p, v=p, w=1-q \\
(0,0,1)\rightarrow(0,0,0) &= 1-p-p-(1-q)+\cC(p,p,1)+\cC(p,1,1-q)+\cC(1,p,1-q)-\cC(p,p,1-q)\\
(0,0,1)\rightarrow(0,0,1) &= (1-q)-\cC(p,1,1-q)-\cC(1,p,1-q)+\cC(p,p,1-q)\\
(0,0,1)\rightarrow(0,1,0) &= p-\cC(p,p,1)-\cC(1,p,1-q)+\cC(p,p,1-q)\\
(0,0,1)\rightarrow(1,0,0) &= p-\cC(p,p,1)-\cC(p,1,1-q)+\cC(p,p,1-q)\\
(0,0,1)\rightarrow(0,1,1) &= \cC(1,p,1-q)-\cC(p,p,1-q) \\
(0,0,1)\rightarrow(1,0,1) &= \cC(p,1,1-q)-\cC(p,p,1-q) \\
(0,0,1)\rightarrow(1,1,0) &= \cC(p,p,1)-\cC(p,p,1-q) \\
(0,0,1)\rightarrow(1,1,1) &= \cC(p,p,1-q) \\
\text{From state} \ (0,1,0), \ &u=p, v=1-q, w=p \\
 (0,1,0)\rightarrow(0,0,0) &= 1-p-(1-q)-p+\cC(p,1-q,1)+\cC(p,1,p)+\cC(1,1-q,p)-\cC(p,1-q,p)\\
(0,1,0)\rightarrow(0,0,1) &= p-\cC(p,1,p)-\cC(1,1-q,p)+\cC(p,1-q,p)\\
(0,1,0)\rightarrow(0,1,0) &= (1-q)-\cC(p,1-q,1)-\cC(1,1-q,p)+\cC(p,1-q,p)\\
(0,1,0)\rightarrow(1,0,0) &= p-\cC(p,1-q,1)-\cC(p,1,p)+\cC(p,1-q,p)\\
(0,1,0)\rightarrow(0,1,1) &= \cC(1,1-q,p)-\cC(p,1-q,p) \\
(0,1,0)\rightarrow(1,0,1) &= \cC(p,1,p)-\cC(p,1-q,p) \\
(0,1,0)\rightarrow(1,1,0) &= \cC(p,1-q,1)-\cC(p,1-q,p) \\
(0,1,0)\rightarrow(1,1,1) &= \cC(p,1-q,p)\\
\text{From state} \ (1,0,0), \ &u=1-q, v=p, w=p \\
(1,0,0)\rightarrow(0,0,0) &= 1-(1-q)-p-p+\cC(1-q,p,1)+\cC(1-q,1,p)+\cC(p,p,1)-\cC(1-q,p,p)\\
(1,0,0)\rightarrow(0,0,1) &= p-\cC(1-q,1,p)-\cC(p,p,1)+\cC(1-q,p,p)\\
(1,0,0)\rightarrow(0,1,0) &= p-\cC(1-q,p,1)-\cC(p,p,1)+\cC(1-q,p,p)\\
(1,0,0)\rightarrow(1,0,0) &= (1-q)-\cC(1-q,p,1)-\cC(1-q,1,p)+\cC(1-q,p,p)\\
(1,0,0)\rightarrow(0,1,1) &= \cC(p,p,1)-\cC(1-q,p,p) \\
(1,0,0)\rightarrow(1,0,1) &= \cC(1-q,1,p)-\cC(1-q,p,p) \\
(1,0,0)\rightarrow(1,1,0) &= \cC(1-q,p,1)-\cC(1-q,p,p) \\
(1,0,0)\rightarrow(1,1,1) &= \cC(1-q,p,p)\\
\text{From state} \ (0,1,1), \ &u=p, v=1-q, w=1-q \\
(0,1,1)\rightarrow(0,0,0) &= 1-p-(1-q)-(1-q)+\cC(p,1-q,1)+\cC(p,1,1-q)+\cC(1,1-q,1-q)-\cC(p,1-q,1-q)\\
(0,1,1)\rightarrow(0,0,1) &= (1-q)-\cC(p,1,1-q)-\cC(1,1-q,1-q)+\cC(p,1-q,1-q)\\
(0,1,1)\rightarrow(0,1,0) &= (1-q)-\cC(p,1-q,1)-\cC(1,1-q,1-q)+\cC(p,1-q,1-q)\\
(0,1,1)\rightarrow(1,0,0) &= p-\cC(p,1-q,1)-\cC(p,1,1-q)+\cC(p,1-q,1-q)\\
(0,1,1)\rightarrow(0,1,1) &= \cC(1,1-q,1-q)-\cC(p,1-q,1-q) \\
(0,1,1)\rightarrow(1,0,1) &= \cC(p,1,1-q)-\cC(p,1-q,1-q) \\
(0,1,1)\rightarrow(1,1,0) &= \cC(p,1-q,1)-\cC(p,1-q,1-q) \\
(0,1,1)\rightarrow(1,1,1) &= \cC(p,1-q,1-q)
\end{align*}
\begin{align*}
    \text{From state} \ (1,0,1), \ &u=1-q, v=p, w=1-q \\
(1,0,1)\rightarrow(0,0,0) &= 1-(1-q)-p-(1-q)+\cC(1-q,p,1)+\cC(1-q,1,1-q)+\cC(p,1,1-q)-\cC(1-q,p,1-q)\\
(1,0,1)\rightarrow(0,0,1) &= (1-q)-\cC(1-q,1,1-q)-\cC(p,1,1-q)+\cC(1-q,p,1-q)\\
(1,0,1)\rightarrow(0,1,0) &= p-\cC(1-q,p,1)-\cC(p,1,1-q)+\cC(1-q,p,1-q)\\
(1,0,1)\rightarrow(1,0,0) &= (1-q)-\cC(1-q,p,1)-\cC(1-q,1,1-q)+\cC(1-q,p,1-q)\\
(1,0,1)\rightarrow(0,1,1) &= \cC(p,1,1-q)-\cC(1-q,p,1-q) \\
(1,0,1)\rightarrow(1,0,1) &= \cC(1-q,1,1-q)-\cC(1-q,p,1-q) \\
(1,0,1)\rightarrow(1,1,0) &= \cC(1-q,p,1)-\cC(1-q,p,1-q) \\
(1,0,1)\rightarrow(1,1,1) &= \cC(1-q,p,1-q)\\
    \text{From state} \ (1,1,0), \ &u=1-q, v=1-q, w=p \\
(1,1,0)\rightarrow(0,0,0) &= 1-(1-q)-(1-q)-p+\cC(1-q,1-q,1)+\cC(1-q,1,p)+\cC(1-q,p,1)-\cC(1-q,1-q,p)\\
(1,1,0)\rightarrow(0,0,1) &= p-\cC(1-q,1,p)-\cC(1-q,p,1)+\cC(1-q,1-q,p)\\
(1,1,0)\rightarrow(0,1,0) &= (1-q)-\cC(1-q,1-q,1)-\cC(1-q,1,p)+\cC(1-q,1-q,p)\\
(1,1,0)\rightarrow(1,0,0) &= (1-q)-\cC(1-q,1-q,1)-\cC(1-q,p,1)+\cC(1-q,1-q,p)\\
(1,1,0)\rightarrow(0,1,1) &= \cC(1-q,p,1)-\cC(1-q,1-q,p) \\
(1,1,0)\rightarrow(1,0,1) &= \cC(1-q,1,p)-\cC(1-q,1-q,p) \\
(1,1,0)\rightarrow(1,1,0) &= \cC(1-q,1-q,1)-\cC(1-q,1-q,p) \\
(1,1,0)\rightarrow(1,1,1) &= \cC(1-q,1-q,p)\\
    \text{From state} \ (1,1,1), \ &u=1-q, v=1-q, w=1-q \\
(1,1,1)\rightarrow(0,0,0) &= 1-(1-q)-(1-q)-(1-q)+\cC(1-q,1-q,1)+\cC(1-q,1,1-q)+\\&\hspace*{1cm}\cC(1,1-q,1-q)-\cC(1-q,1-q,1-q)\\
(1,1,1)\rightarrow(0,0,1) &= (1-q)-\cC(1-q,1,1-q)-\cC(1-q,1-q,1)+\cC(1-q,1-q,1-q)\\
(1,1,1)\rightarrow(0,1,0) &= (1-q)-\cC(1-q,1-q,1)-\cC(1-q,1,1-q)+\cC(1-q,1-q,1-q)\\
(1,1,1)\rightarrow(1,0,0) &= (1-q)-\cC(1-q,1-q,1)-\cC(1-q,1,1-q)+\cC(1-q,1-q,1-q)\\
(1,1,1)\rightarrow(0,1,1) &= \cC(1-q,1-q,1)-\cC(1-q,1-q,1-q) \\
(1,1,1)\rightarrow(1,0,1) &= \cC(1-q,1,1-q)-\cC(1-q,1-q,1-q) \\
(1,1,1)\rightarrow(1,1,0) &= \cC(1-q,1-q,1)-\cC(1-q,1-q,1-q) \\
(1,1,1)\rightarrow(1,1,1) &= \cC(1-q,1-q,1-q).
\end{align*}

\subsection{Gradients and Hessians from the log likelihood and log priors}
\label{sec:GradientsAppendix}
% The complete-data log likelihood is decomposed as:
% \begin{align*}
%     \text{log} \ \prob(\by, \bx|\btheta) &= \text{log} \ \prob(\by|\bx,\btheta) + \text{log} \ \prob(\bx|\btheta)
% \end{align*}
% and then the Fisher identity implies:
% \begin{align*}
%    \nabla_{\btheta}\text{log} \ \prob(\by|\btheta) &= \mathbb{E}_{\bx|\by,\btheta}\left[\nabla_{\btheta}\text{log} \ \prob(\by|\bx,\btheta)\right] + \mathbb{E}_{\bx|\by,\btheta} \left[\nabla_{\btheta}\text{log}\prob(\bx|\btheta)\right].
% \end{align*}
% Similarly, our chosen metric tensor for the log likelihood implies:
% \begin{align*}
%     \nabla_{\btheta}^2 \ \text{log} \ \prob(\by|\btheta) &\approx \sum_{\bx \in \cX} \nabla_{\btheta}^2 \ \text{log} \ \prob(\by, \bx|\btheta)\prob(\bx|\by,\btheta)=\mathbb{E}_{\bx|\by,\btheta}\left[\nabla_{\btheta}^2\text{log} \ \prob(\by|\bx,\btheta)\right] + \mathbb{E}_{\bx|\by,\btheta} \left[\nabla_{\btheta}^2\text{log}\prob(\bx|\btheta)\right].
% \end{align*}
% We emphasize that the log prior for the joint-state sequence $\text{log} \ \prob(\bx|\btheta)$ in Equation \ref{eq:complete-data loglikelihood decomposition} does not contribute to the gradients and Hessians of the log marginal likelihood with respect to the temporal and spatial components ($\boldsymbol{r}, \boldsymbol{s},\boldsymbol{u}$) since it only depends on $\bGamma \in \btheta$. 

% Recall the general form of our model in Equation \ref{eq:model-multitype}. The conditional log likelihood given the joint-state sequence in the context of our model is written as:
The gradients and Hessians from the log likelihood with respect to $\br,\bs$ and $\bu$ are:
 \begin{align*}
%     \text{log} \ \prob(\by_{1:I,1:T,1:K}|\bx,\btheta) &= \sum_{i=1}^{I}\sum_{t=1}^T\sum_{k=1}^K (y_{itk}(a_k + r_t + s_{t\bmod C} + u_i + x_{itk}\beta_k) - \exp{(a_k + r_t + s_{t\bmod C} + u_i + x_{itk}\beta_k)}e_{it} \notag \\
%     \quad &+ y_{itk} \ \text{log} \ e_{it} - \text{log} \ y_{itk}!). 
% \end{align*}
% We now define $\eta_{itk}$ as the following expectation, where we smooth $e_{it}\lambda_{itk}$ and marginalize over the state space $\cX$:
% \begin{align*}
%     \eta_{itk}&:=\mathbb{E}_{\bx|y_{1:I,1:T,1:K},\btheta}\left[e_{it}\lambda_{itk}\right] = \sum_{\bx_{it}\in \cX}e_{it}\exp(a_k + r_t + s_{t\bmod C} + u_i + x_{itk}\beta_k) \prob(\bx_{it}|y_{i,1:T,1:K},\btheta) \label{eq:smoothing}, 
% \end{align*}
% where $\prob(\bx_{it}|y_{i,1:T,1:K},\btheta)$ is obtained as described in Section \ref{sec:outBprob}. 
% The gradients and approximate Hessians for the temporal and spatial components from the incomplete-data log likelihood are given below:
% \begin{align*}
% \nabla_{r_t} \ \text{log} \ \prob(\by_{1:I,1:T,1:K}|\btheta) &= \sum_{i=1}^{I}\sum_{k=1}^K \left(y_{itk} - \eta_{itk} \right) \\
    \nabla_{\boldsymbol{r}} \ \text{log} \ \prob(\by_{1:I,1:T,1:K}|\btheta) &= \left[\sum_{i=1}^{I}\sum_{k=1}^K \left(y_{i1k} - \eta_{i1k} \right),\dots,\sum_{i=1}^{I}\sum_{k=1}^K \left(y_{iTk} - \eta_{iTk} \right)\right]^\top \\
     % \nabla_{r_t}^2 \ \text{log} \ \prob(\by_{1:I,1:T,1:K}|\btheta) &\approx \sum_{i=1}^{I}\sum_{k=1}^K \left(- \eta_{itk} \right) \\
     \nabla_{\boldsymbol{r}}^2 \ \text{log} \ \prob(\by_{1:I,1:T,1:K}|\btheta) &\approx \text{diag} \left[\sum_{i=1}^{I}\sum_{k=1}^K \left(- \eta_{i1k} \right),\dots,\sum_{i=1}^{I}\sum_{k=1}^K \left(- \eta_{iTk} \right) \right] \\
    % \nabla_{s_{t\bmod C}} \ \text{log} \ \prob(\by_{1:I,1:T,1:K}|\btheta) &= \sum_{t:t\bmod C=c}\sum_{i=1}^{I}\sum_{k=1}^K \left(y_{itk} - \eta_{itk} \right) \\ 
    \nabla_{\boldsymbol{s}} \ \text{log} \ \prob(\by_{1:I,1:T,1:K}|\btheta) &= \left[\sum_{t:t\bmod C=1}\sum_{i=1}^{I}\sum_{k=1}^K \left(y_{itk} - \eta_{itk} \right),\dots,\sum_{t:t\bmod C=C}\sum_{i=1}^{I}\sum_{k=1}^K \left(y_{itk} - \eta_{itk} \right) \right]^\top \\ 
    % \nabla_{s_{t \bmod C}}^2 \ \text{log} \ \prob(\by_{1:I,1:T,1:K}|\btheta) &\approx \sum_{t:t\bmod C = c}\sum_{i=1}^{I}\sum_{k=1}^K \left(- \eta_{itk} \right) \\ 
    \nabla_{\boldsymbol{s}}^2 \ \text{log} \ \prob(\by_{1:I,1:T,1:K}|\btheta) &\approx \text{diag}\left[\sum_{t:t\bmod C = 1}\sum_{i=1}^{I}\sum_{k=1}^K \left(- \eta_{itk} \right),\dots,\sum_{t:t\bmod C = C}\sum_{i=1}^{I}\sum_{k=1}^K \left(- \eta_{itk} \right) \right] \\ 
    % \nabla_{u_i} \ \text{log} \ \prob(\by_{1:I,1:T,1:K}|\btheta) &= \sum_{t=1}^{T}\sum_{k=1}^K \left(y_{itk} - \eta_{itk} \right) \\ 
    % \nabla_{u_i}^2 \ \text{log} \ \prob(\by_{1:I,1:T,1:K}|\btheta) &\approx \sum_{t=1}^{T}\sum_{k=1}^K \left(- \eta_{itk} \right)\\
    \nabla_{\boldsymbol{u}} \ \text{log} \ \prob(\by_{1:I,1:T,1:K}|\btheta) &= \left[\sum_{t=1}^{T}\sum_{k=1}^K \left(y_{1tk} - \eta_{1tk} \right),\dots,\sum_{t=1}^{T}\sum_{k=1}^K \left(y_{Itk} - \eta_{Itk} \right) \right]^\top\\ 
    \nabla_{\boldsymbol{u}}^2 \ \text{log} \ \prob(\by_{1:I,1:T,1:K}|\btheta) &\approx \text{diag}\left[\sum_{t=1}^{T}\sum_{k=1}^K \left(- \eta_{1tk} \right),\dots,\sum_{t=1}^{T}\sum_{k=1}^K \left(- \eta_{Itk} \right) \right]
%     \intertext{The cross-derivatives in the Hessians of $\text{log} \ \prob(\by|\bx,\btheta)$ with respect to $\br,\bs, \text{and} \ \bu$ are all zeros since these components at a given time point $t$ make no contribution to the log likelihood at any other time point.}
\end{align*}
The log prior densities for the temporal and spatial components are given by:
\begin{align*}
    \text{log} \ \prob(\boldsymbol{r}|\kappa_r) &= -\frac{\kappa_r}{2}\boldsymbol{r}^\top\boldsymbol{R}_r\boldsymbol{r}+\const \\
    \text{log} \ \prob(\boldsymbol{s}|\kappa_s) &= -\frac{\kappa_s}{2}\boldsymbol{s}^\top\boldsymbol{R}_c\boldsymbol{s}+\const \\
    \text{log} \ \prob(\boldsymbol{u}|\kappa_u) &= -\frac{\kappa_u}{2}\boldsymbol{u}^\top\boldsymbol{R}_u\boldsymbol{u}+\const \\
\end{align*}
The gradients and Hessians for these log prior densities with respect to $\br,\bs$ and $\bu$ are:
\begin{align*}
    \nabla_{\boldsymbol{r}} \ \text{log} \ \prob(\boldsymbol{r}|\kappa_r) &= \frac{-\kappa_r(\boldsymbol{R}_r+\boldsymbol{R}_r^\top)\boldsymbol{r}}{2}=-\kappa_r\boldsymbol{R}_r\boldsymbol{r}\ \\
 \nabla_{\boldsymbol{s}} \ \text{log} \ \prob(\boldsymbol{s}|\kappa_s) &= \frac{-\kappa_s(\boldsymbol{R}_c+\boldsymbol{R}_c^\top)\boldsymbol{s}}{2}= -\kappa_s\boldsymbol{R}_c\boldsymbol{s} \\
   \nabla_{\boldsymbol{u}} \ \text{log} \ \prob(\boldsymbol{u}|\kappa_u) &=\frac{-\kappa_u(\boldsymbol{R}_u+\boldsymbol{R}_u^\top)\boldsymbol{u}}{2}= -\kappa_u\boldsymbol{R}_u\boldsymbol{u} \\ 
        \nabla_{\boldsymbol{r}}^2 \ \text{log} \ \prob(\boldsymbol{r}|\kappa_r) &= -\kappa_r\boldsymbol{R_r} \\
     \nabla_{\boldsymbol{s}}^2 \ \text{log} \ \prob(\boldsymbol{s}|\kappa_s) &= -\kappa_s\boldsymbol{R}_c \\
     \nabla_{\boldsymbol{u}}^2 \ \text{log} \ \prob(\boldsymbol{u}|\kappa_u) &= -\kappa_u\boldsymbol{R}_u
\end{align*}

\clearpage
\bibliographystyle{unsrt}
\bibliography{Bibliography}  

\newpage
\pagenumbering{gobble}
\setcounter{figure}{0}
\setcounter{table}{0}
\makeatletter 
\renewcommand{\thefigure}{S\@arabic\c@figure} 
\renewcommand{\thetable}{S\@arabic\c@table} 
\makeatother

\section*{Supplementary Material}

\begin{figure}[H]
\centerline{\includegraphics[width=6.5in, height=3.5in, keepaspectratio=true]{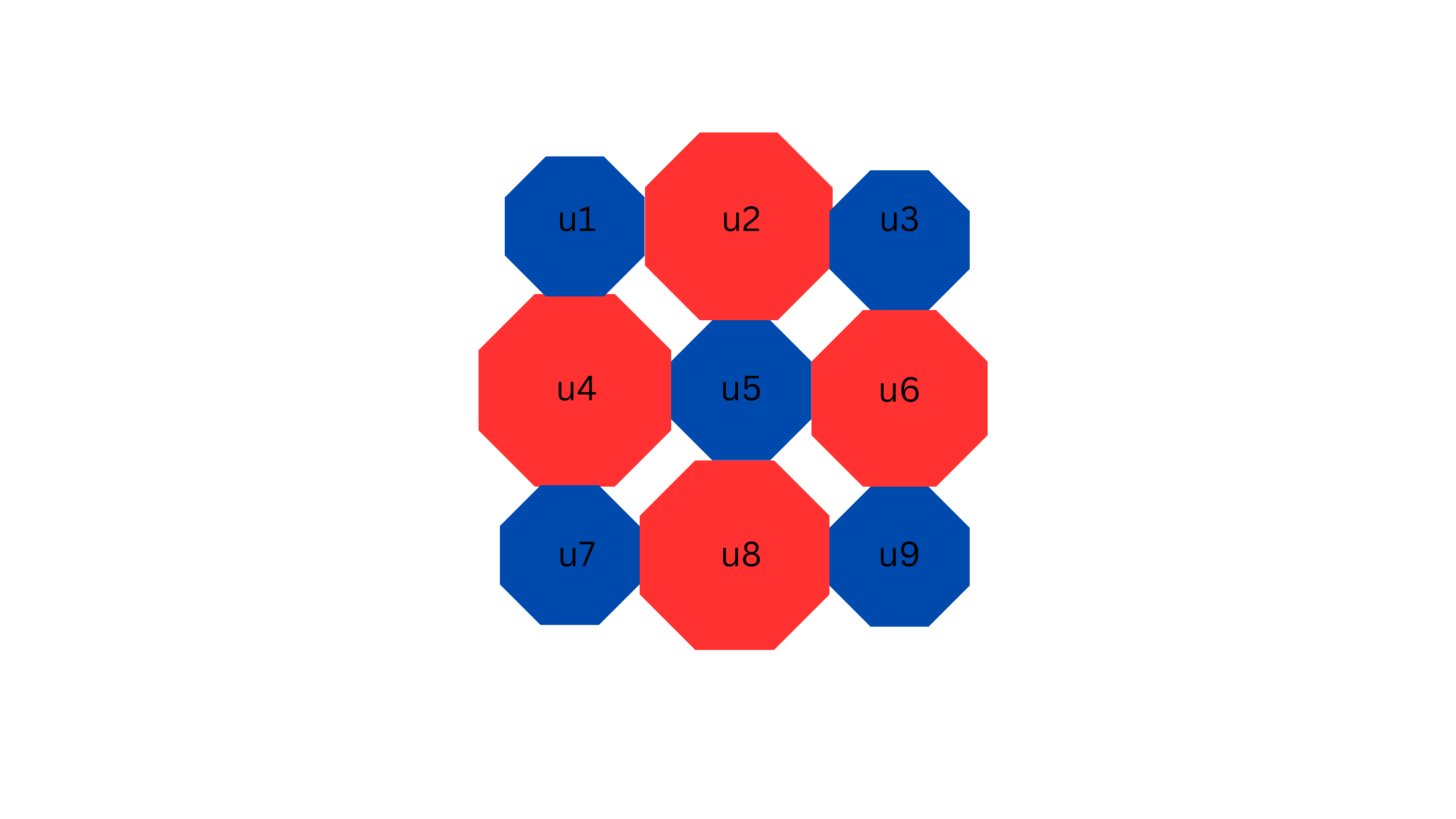}}
\caption{Adjacency structure of spatial locations with small cities in blue and large cities in red.}
\label{fig:simulationCities}
\end{figure}

\begin{figure}[H]
\centerline{\includegraphics[width=1\textwidth]{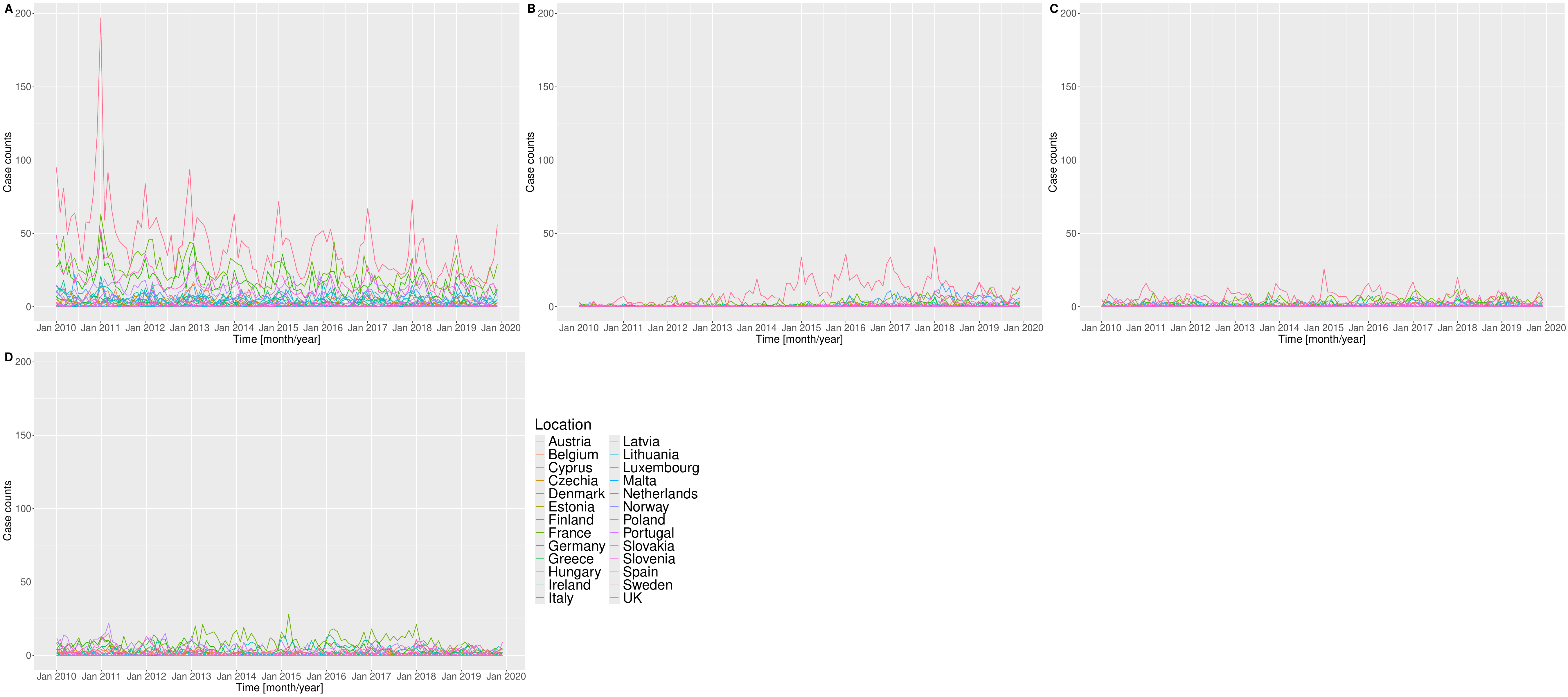}}
\caption{Monthly reported case counts of multi-type invasive meningococcal disease in 26 European countries from January 2010 to December 2019. The disease types are (A) NEIMENI-B, (B) NEIMENI-W, (C) NEIMENI-Y, and (D) NEIMENI-C.}
\label{fig:applicationdata}
\end{figure}

\begin{figure}[H]
\centerline{\includegraphics[width=\textwidth, height=\textheight, keepaspectratio]{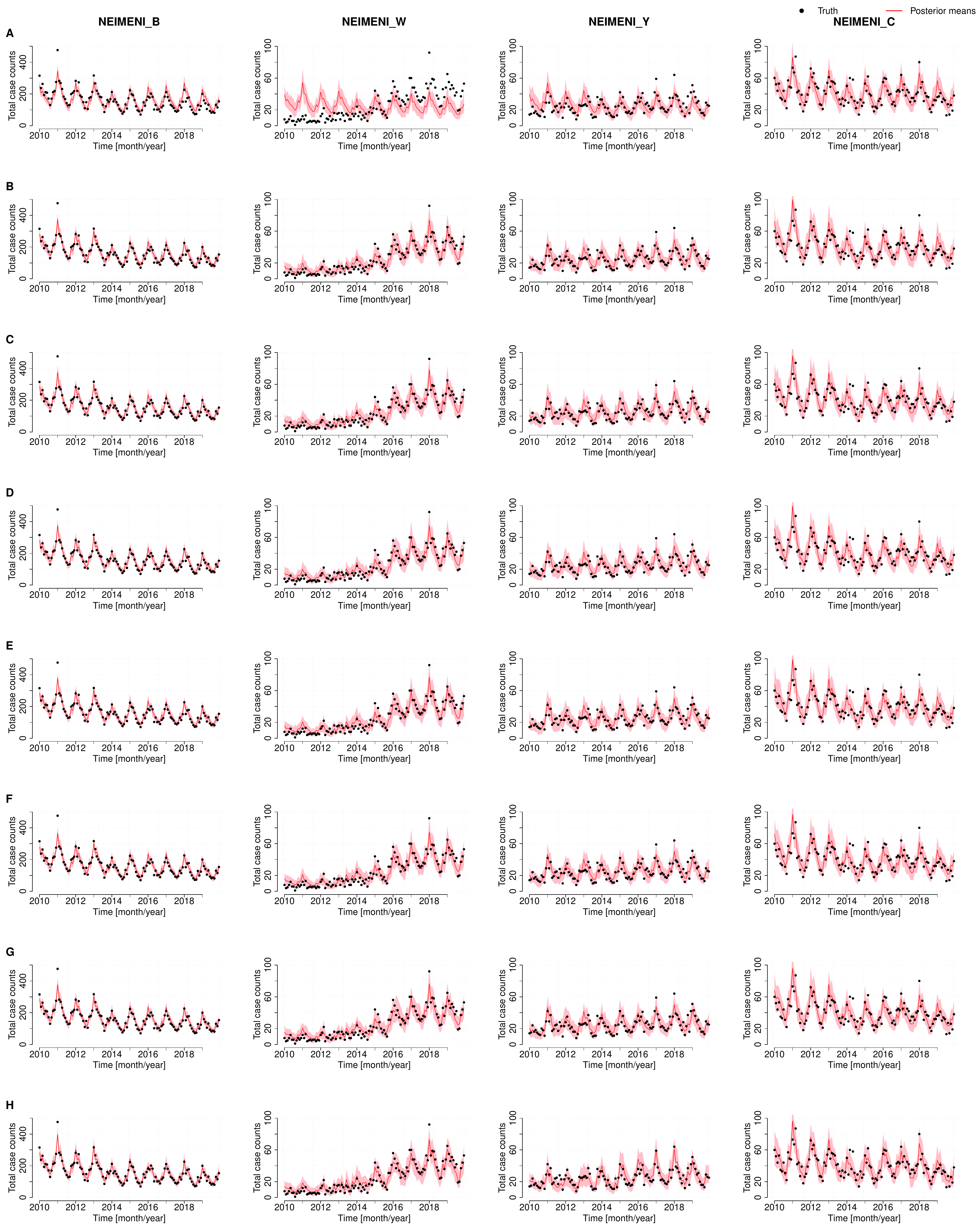}}
\caption{Posterior predictive fits for the multitype meningococcal disease analysis for the overall case counts in each serogroup across all spatial locations for (A) No epidemic model (B) Independent 1 model (C) Independent 2 model (D) Frank copula 1 model (F) Frank copula 2 model (G) Gaussian factor copula 1 model (H) Gaussian factor copula 2 model, and (H) General-dependent model. Real data shown with black dots.}
\label{fig:applicationposteriorpredictives}
\end{figure}

\end{document}